\pdfoutput=1
\documentclass[journal]{IEEEtran}

\usepackage[utf8]{inputenc}
\usepackage{graphics,epsfig,epstopdf}
\usepackage{url,amsmath,amssymb}
\usepackage{array}
\usepackage{color,subfigure}
\usepackage{booktabs}
\usepackage{threeparttable}
\usepackage{multirow}
\usepackage{tabularx}
\usepackage{verbatim}
\usepackage[table]{xcolor}
\usepackage{hyperref} 
\usepackage{booktabs}
\usepackage[noend]{algorithm}  
\usepackage[noend]{algpseudocode}  
\usepackage[numbers]{natbib}

\usepackage{graphicx}
\usepackage{float}
\usepackage{enumitem}
\usepackage{subfigure}
\begin{document}
\bibliographystyle{IEEEtran}

\title{INNOVATIVE SEMANTIC COMMUNICATION SYSTEM}

\author{Chen Dong,
        Haotai Liang*,
        Xiaodong Xu,~\IEEEmembership{Senior Member,~IEEE,}
        Shujun Han, 
        Bizhu Wang,
        Ping Zhang,~\IEEEmembership{Fellow,~IEEE}

\thanks{Chen Dong, Xiaodong Xu, Shujun Han, Bizhu Wang and Ping Zhang are with the State Key Laboratory of Networking and Switching Technology, Beijing University of Posts and Telecommunications, Beijing, China (e-mail: dongchen@bupt.edu.cn; xuxiaodong@bupt.edu.cn; hanshujun@bupt.edu.cn; wangbizhu\_7@bupt.edu.cn;  pzhang@bupt.edu.cn).}

\thanks{*Haotai Liang is the corresponding author and with the School of Electronics and Information Engineering, Shenzhen University, Shenzhen, China (e-mail: 2018133021@email.szu.edu.cn).}
}

\markboth{Journal of \LaTeX\ Class Files,~Vol.~xx, No.~x, xx~xxx}%
{Shell \MakeLowercase{\textit{et al.}}: Bare Demo of IEEEtran.cls for IEEE Journals}

\maketitle

\begin{abstract}
Traditional communication systems focus on the transmission process, and the context-dependent meaning has been ignored. The fact that 5G system has approached Shannon limit and the increasing amount of data will cause communication bottleneck, such as the increased delay problems. Inspired by the ability of artificial intelligence to understand semantics, we propose a new communication paradigm, which integrates artificial intelligence and communication, the semantic communication system.  Semantic communication is at the second level of communication based on Shannon and Weaver\cite{6197583}, which retains the semantic features of the transmitted information and recovers the signal at the receiver, thus compressing the communication traffic without losing important information. Different from other semantic communication systems, the proposed system not only transmits semantic information but also transmits semantic decoder. In addition, a general semantic metrics is proposed to measure the quality of semantic communication system. In particular, the semantic communication system for image, namely AESC-I, is designed to verify the feasibility of the new paradigm. Simulations are conducted on our system with the additive white Gaussian noise (AWGN) and the multipath fading channel using MNIST and Cifar10 datasets. The experimental results show that DeepSC-I can effectively extract semantic information and reconstruct images at a relatively low SNR.

\end{abstract}


\IEEEpeerreviewmaketitle

\section{Introduction}
\IEEEPARstart{W}{ith} the advent of the fifth generation(5G), the achieved transmission rate has been increased by thousands of times compared with that in the fourth generation and the achieved transmission rate has once again approached the Shannon limit, but there is no limit to the growth of data. In communication system that only pays attention to symbol efficiency and accuracy measured by bits, with limited bandwidth resources but low latency requirements, the existing source channel coding technology suffers great challenge facing the continuous explosion of data. The semantic communication achieves the goal of extremely compressing the communication traffic by filtering redundant, irrelevant, unconcerned information and extracting the meaning of effective information~\cite{9475174, ZHANG2021}.
	
Inspired by the semantic level communication based on Shannon and Weaver~\cite{6197583}, the semantic information sent by the transmitter and the meaning of the semantic information interpreted by the receiver. In addition, the ability of artificial intelligence to understand human information in limited scenarios is gradually increasing. Deep learning has achieved great success in understanding language, voice, image and other information sources as one of the most important technologies of artificial intelligence. In terms of natural language processing (NLP), neural network methods usually use low-dimensional and dense vectors to implicitly express the semantic features of the language. At present, the pre-training model (PTM) trained by a large corpus~\cite{mikolov2013distributed, devlin2018bert, mccann2017learned, sarzynska2021detecting} has been applied in various NLP task. The pre-trained embedding code captures the high-level meaning of words and sentences in the upstream of the NLP task, avoiding the downstream task to understand the language on the symbols from the beginning. When it comes to image semantic analysis, most of the deep learning models \cite{he2016deep, ren2015faster, isensee2019nnu, goodfellow2014generative, karras2019style, dahl2011context} directly output the results of semantic analysis, such as target recognition, target detection, segmentation and scene understanding, etc. These tasks can correctly assign semantic tags. But it is worth mentioning that the current emergence of Generative Adversarial Neural Networks(GAN)\cite{goodfellow2014generative} has deepened the interpretability of image semantics. For example, StyleGan\cite{karras2019style} can control different attributes to generate corresponding style images. The most typical applications of speech semantics are speech recognition human interaction systems, speech-to-text systems. Microsoft’s context-dependent DNN-HMM research\cite{dahl2011context} results on large vocabulary speech recognition have completely changed the original technical framework of speech recognition system, which doesn’t care about the speaking speed and tone, but only extracts the text of the speech. 

At the same time, deep learning has not only made breakthroughs in natural language processing, speech recognition, image processing, communication systems based on deep learning frameworks have also been developed in recent years, showing great potential for its performance to surpass traditional communication systems. Authors in \cite{o2016learning} proposed as communication system based on the end-to-end learning and optimization of autoencoders and its pioneering work provides inspiration for the design of DL-based communication systems.
	
Thanks to the above enlightenment and the development of deep learning, semantic communication has become a breakthrough to improve communication efficiency again. A series of semantic communication systems based on deep learning are proposed. Authors in \cite{xie2021deep} propose a deep learning based semantic communication system, named DeepSC, for text transmission, which aims at maximizing the system capacity and minimizing the semantic errors by recovering the meaning of sentences, rather than bit or symbol errors. To make the DeepSC affordable for Internet-of-Things devices, authors in \cite{xie2020lite} propose a model compression algorithm, including network sparsification and quantization, to reduce the size of DL models by pruning the redundancy connections and quantizing the weights. Similarly, a DL-enabled semantic communication system for speech signals is first developed in \cite{weng2021semantic}, which can learn and extract speech signals and then recover them at the receiver from the received features directly.

The current semantic model based on deep learning is aimed at the certain information sources. Considering the actual communication process, the receiving node is required to have a decoder of all information sources, which undoubtedly increases the load and power consumption of the node. Xie and Qin proposed a distributed semantic communication system including three steps in \cite{xie2020lite}: Model Initialization/Update, Model Broadcasting, Semantic Features Upload. The method of propagating semantic model parameters allows nodes to have the ability to receive any information source, but in the face of new information sources, the paper does not talk about how to update the initialization model of the cloud platform. In addition, the investigation on semantic communication for image signals transmission is still missed, but there is also an image transmission system that reflects the spirit of sharing semantic information, such as an image retrieval application \cite{9066966} based on Joint Source Channel Coding (JSCC) model, which aims to reduce the transmission delay of Internet of Things devices by implementing image compression facing retrieval. The JSCC model is also used to perform image classification tasks on edge servers ~\cite{9154306}, helping the Internet of Things process images with lower computational complexity and reduce transmission bandwidth.

In this paper, we consider a new communication paradigm and design a semantic communication system for image signals based on autoencoder, named AESC-I, which can extract the semantic features of the image and restore the source image at the receiver.  The main contributions of this paper are summarized as follows:\begin{itemize} 
\item A innovative semantic communication paradigm integrating artificial intelligence(AI) communication is proposed. The semantic communication system is designed on the basis of Shannon-based physical layer, and the AI model is integrated on the transmitter and receiver. The transmitter not only transmits semantic coding, but also needs to send the AI-based decoder together.
\item From the practical meaning of semantics, combined with the proposed semantic communication system, a general semantic metrics is proposed.
\item An image semantic communication system based on the autoencoder for verification is designed. Inspired by the semantic meaning, we propose a loss function which can effectively extract the semantic features of images, and apply it to the training of semantic encoder and decoder. 
\end{itemize}	

This paper is arranged as follows: Section 2 introduces semantic communication system and the performance metrics. Section 3 details the proposed semantic communication system for image. Numerical results are presented in section 4. Finally, conclusions about out work are drawn in section 5.

\section{System model and performance metrics}

\begin{figure*}
    \centering
    \setlength{\abovecaptionskip}{0.cm}
    \includegraphics[width=\textwidth]{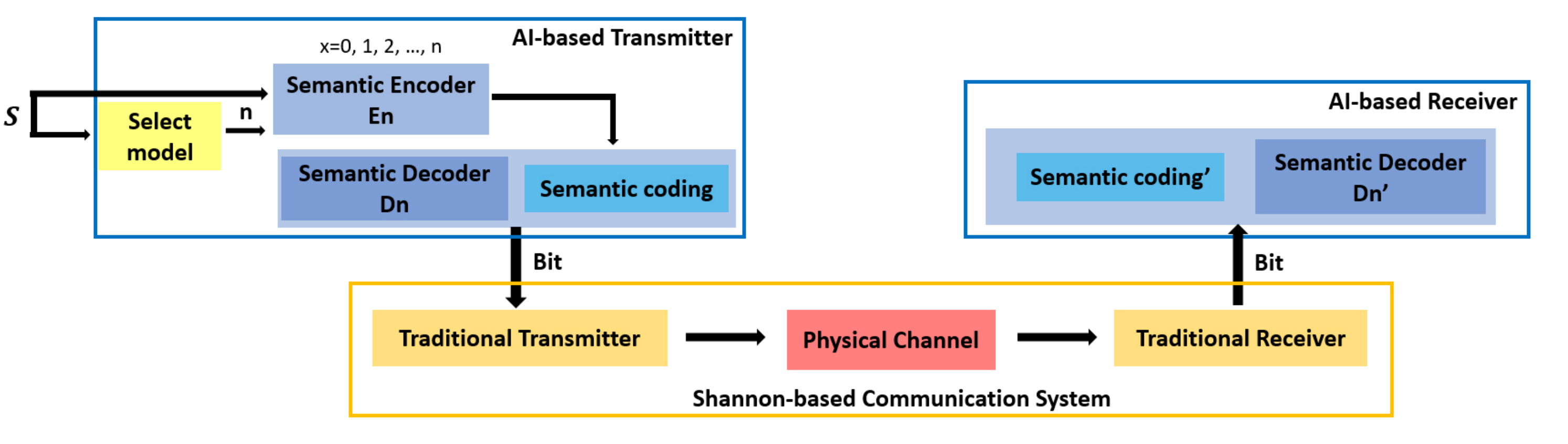}
    \centering
    \caption{Architecture of the innovative semantic communication system. The system architecture consists of AI-based transmitter, receiver and Shannon-based communication systems, where the semantic decoder represented by purple blocks and the semantic coding represented by blue blocks pass through the physical channel as bit streams, and the receiver uses the received semantic decoder to recover the source.}
    \label{semantic system}
\end{figure*}

The architecture of the semantic communication system is shown in Fig.\ref{semantic system}. Different from the existing semantic communication system architecture proposed in \cite{xie2021deep}, \cite{xie2020lite}, and \cite{weng2021semantic}, it is based on the traditional Shannon communication system and does not affect each module of Shannon physical layer. However, inspired by their semantic communication architecture, we embed the AI model into the transmitter and receiver of the physical layer to perform effective semantic communication. It is worth noting that we did not deliberately reduce the workload of nodes by compressing the AI model. Instead of deploying the AI model in advance at all nodes, we first pre-train the semantic encoder and semantic decoder at the transmitter. According to the requirement of the upper level, we transmitted the selected decoder and semantic coding encoded by the selected encoder, and the receiver recovered the received semantic coding through the received decoder to achieve the effect of semantic communication.

At the source side, it is necessary to identify the category of the information source, select the AI model matching the demand according to the classification or clustering algorithm. The model will extract and compress the semantics of the source, and then package the semantic coding with the AI model to form a bit stream for Shannon-based communication system. The information received by the receiver is the integration of semantic information, AI model and environment information. Environmental information, such as spectrum environment and electromagnetic environment, can be interpreted through traditional physical layer. Shannon-based channel will solve the work of channel  encoder and modulation, and the processing level is bit and symbol. The level of the AI-based semantic layer processing is the meaning of source information. 

Model is the key to semantic features. The transmission and updating of model also means the continuous iteration of semantic features. The essence of the AI model is the abstraction of data, and the continuous growth of data means that the AI model is not invariable, and the function of semantic feature extraction of the AI model needs to be updated constantly. The integration of AI model propagation in semantic communication system can effectively and continuously update the AI model. Subsequently, it can also be updated according to the differences of AI models between different nodes, so as to reduce the amount of the data transmitted. 

Due to the existence of model propagation, the proposed communication system does not rely on the pre-trained decoder like \cite{xie2021deep}\cite{weng2021semantic}, and can receive and integrate the AI models from other nodes according to different conditions such as service requirements and quality. Therefore, in terms of performance evaluation, the transmission of semantic layer should pay attention to whether the information recovered by semantics can meet the expectations of subsequent tasks. A general quality index of semantic service is given by,

\begin{equation}\label{metric}
    SS = \frac{ST(\hat{S})}{ST(S)},
\end{equation}
where $S$ denotes the unprocessed information at the transmitter, $\hat{S}$ represents the information recovered through semantics at the receiver and $ST(\cdot)$ represents the performance of the source in performing subsequent tasks. We suggest using $Sigmoid$ and other functions to map the result of $ST(\cdot)$ to $[0, 1]$, and the value of $Sigmoid(ST(S))$ should be close to 1. The indicator for selecting subsequent tasks should be that the better the performance, the higher the indicator. If $SS$ is equal to 1, the semantic information is completely recovered. If $SS$ is equal to 0, it fails to recovered.

\section{Semantic Communication System for Image}
In this section, the design and training methods of the proposed semantic communication system for image based on AutoEncoder namely AESC-I is introduced, where the system diagram is illustrated in Fig.\ref{Image system}. Specifically, the developed autoencoder based semantic encoder/decoder will be detailed as shown in experiment section. The designed semantic communication system for image is different from the JSCC in \cite{8723589}. They only consider the structural image source, and add the function of the joint source-channel coding to the model. However, we need to transmit not only structural images, but also non structural model parameters. We focus on semantic transmission by using AI model based on Shannon physical layer.

\subsection{Problem formulation}
After identifying that the source is the image source by classification or clustering, the transmitter selects the appropriate semantic encoder and decoder. We denote the image source by $S \in R^{w\cdot h\cdot c}$, where $w$,$h$,$c$  represent the width, height, and the number of channels of the image respectively. The semantic encoder composed of the convolutional downsampling layers and linear layers flattens the image source $S$ into the semantic coding $l \in R^d$, where $d$ represent the dimension of the semantic coding. Denote the parameters of the semantic encoder and decoder as $\alpha$ and $\beta$, respectively. Then the symbols stream, $X$,  is composed of non-structural decoder parameters and structural semantic coding according to certain splicing rules $SR(\cdot)$, given by
\begin{equation}
    X = SR(l, \beta),
\end{equation}

\begin{equation}
    l = E_{\alpha}(S),
\end{equation}
where $E_{\alpha}$ and $D_{\beta}$ indicate the semantic enocder and decoder with respect to parameters $\alpha$ and $\beta$, respectively. The symbol stream is mapped to another symbol stream through the traditional communication process of quantization, channel enocder and modulation.

Then, the multipath fading channel which the symbol stream passes through is modeled as 

\begin{equation}
    y = x \otimes h + n,
\end{equation}
where the vectors $x$ and $y$ represent the transmitted and received symbol stream, respectively, and $n$ represents the white noise. $h$ is channel impulse responses of the multipath channel and $\otimes$ denotes the convolution operation. We assume that the channel response is constant within a symbol stream, and changes between different symbols at the same time. The tapped delay line model of $h$ is given by

\begin{equation}
    h(\tau,t) = \sum_{k=1}^{K}a_k(t)\delta(\tau-\tau_k),
\end{equation}
where $a_k (t)$ represents the channel gain on the $kth$ path, $\tau_k$ is the delay of the $k-th$ path, and $K$ is the total number of paths.

After passing through the channel, the symbol stream is demodulated and channel decoded through a symmetrical process to restore the original symbol stream. The decoder parameters $\beta$ and semantic coding $l$ can be obtained inversely through the previously set rules,
\begin{equation}
    [\hat{\beta}, \hat{l}]=SR^{-1}(y).
\end{equation}
At the receiver, the decoded image can be represented as 
\begin{equation}
    \hat{S} = D_{\hat{\beta}}(\hat{l}).
\end{equation}

The purpose of the semantic communication system is to preserve the semantic of the transmitted signal as much as possible, and the restored signal does not affect subsequent tasks. To make the decoded image and the source image as similar as possible, the mean-squared error (MSE) is used as part of the loss function for training the semantic encoder and decoder, denoted as 

\begin{equation}
    L_{MSE}(S,\hat{S}) = \frac{1}{w \cdot h \cdot  c}\sum_{i=1}^{w}\sum_{j=1}^{h}\sum_{k=1}^{c}(S_{ijk}-{\hat S_{ijk}})^2.
\end{equation}
Considering that the semantic communication system transmits effective semantic features as much as possible, we use the most common classification model in the image field to measure semantic features because effective semantic features can be regarded as that contributes to subsequent task. The semantic errors between the original image and the decoded image can be formulated as

\begin{equation}
L_{SE}(S, \hat{S}) = L_{MSE}(M(S),M(\hat{S})),
\end{equation}
where $M$ is the output of the penultimate layer of the perfect model that can perform classification tasks on image datasets.

The loss function used to train the semantic encoder and decoder is expressed as 

\begin{equation}\label{loss}
L(S,\hat{S}) = \gamma \cdot L_{SE} + (1-\gamma) \cdot L_{MSE},
\end{equation}
where $\gamma \in(0,1)$ weighs the importance of the two loss functions. Generally, $L_{MSE}$ helps the semantic model converge faster.

\subsection{Semantic encoder and decoder}
\begin{figure*}
    \centering
    \setlength{\abovecaptionskip}{0.cm}
    \includegraphics[width=\textwidth]{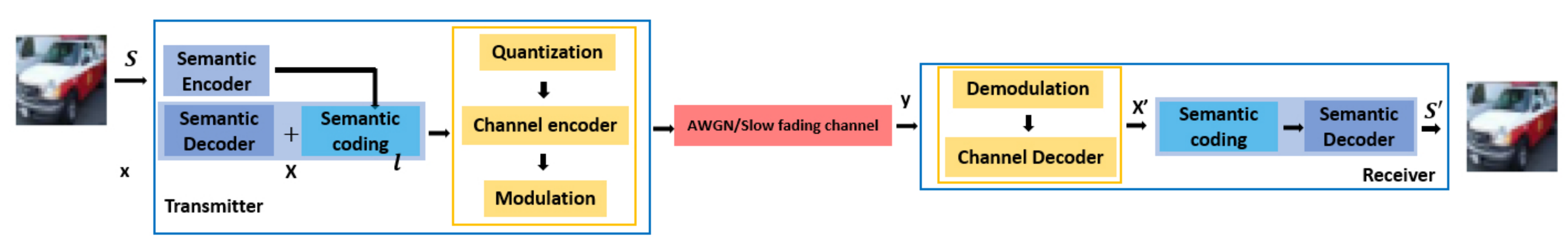}
    \centering
    \caption{Innovative semantic communication for image, AESC-I. Purple blocks represent semantic encoder and decoder, blue blocks represent semantic coding, and yellow blocks represent Shannon-based communication process, including quantization, channel encoder and modulation.}
    \label{Image system}
\end{figure*}

As shown in Fig.\ref{Image system}, the semantic encoder and decoder are both trained and stored in the transmitter. Since the decoder needs to be sent together with the semantic coding, the structure of the decoder needs to be designed to be as small as possible. The basic structure of the convolutional autoencoder (CAE) is used to greatly reduces the redundancy of the parameters compared with the fully connected autoencoder. Because its weights are shared among all position in the image and the spatial locality of the image is preserved. 

For a certain channel input $x$, the sematic latent representation of the $k$th feature map is given by,
\begin{equation}
l^k = {\sigma(x*W^k)}
\end{equation}
where $\sigma$ is an activation function and $*$ denotes the 2D convolution. 

As mentioned above, the convolution operation only extracts local information, and the full connected layer that we added next to it can make the local information to interact with each other. I think it will remove redundancy between the local semantic and other semantic, which achieves the goal of compress semantic information again.

The fully connected layer of the receiver is symmetrical to it and the reconstruction model recovered by the receiver restores the images from the semantic features. The reconstruction is obtained using

\begin{equation}
\hat{x} = {\sigma(\sum_{k\in L}h^k * W^k)},
\end{equation}
where $L$ identifies the group of semantic feature maps, $W$ identifies the flip operation over both dimensions of the weights.

\begin{figure}
    \centering
    \setlength{\abovecaptionskip}{0.cm}
    \subfigure[Training and validation loss of the BCE loss]{
    \begin{minipage}[b]{0.4\textwidth}
    \includegraphics[width=1\textwidth]{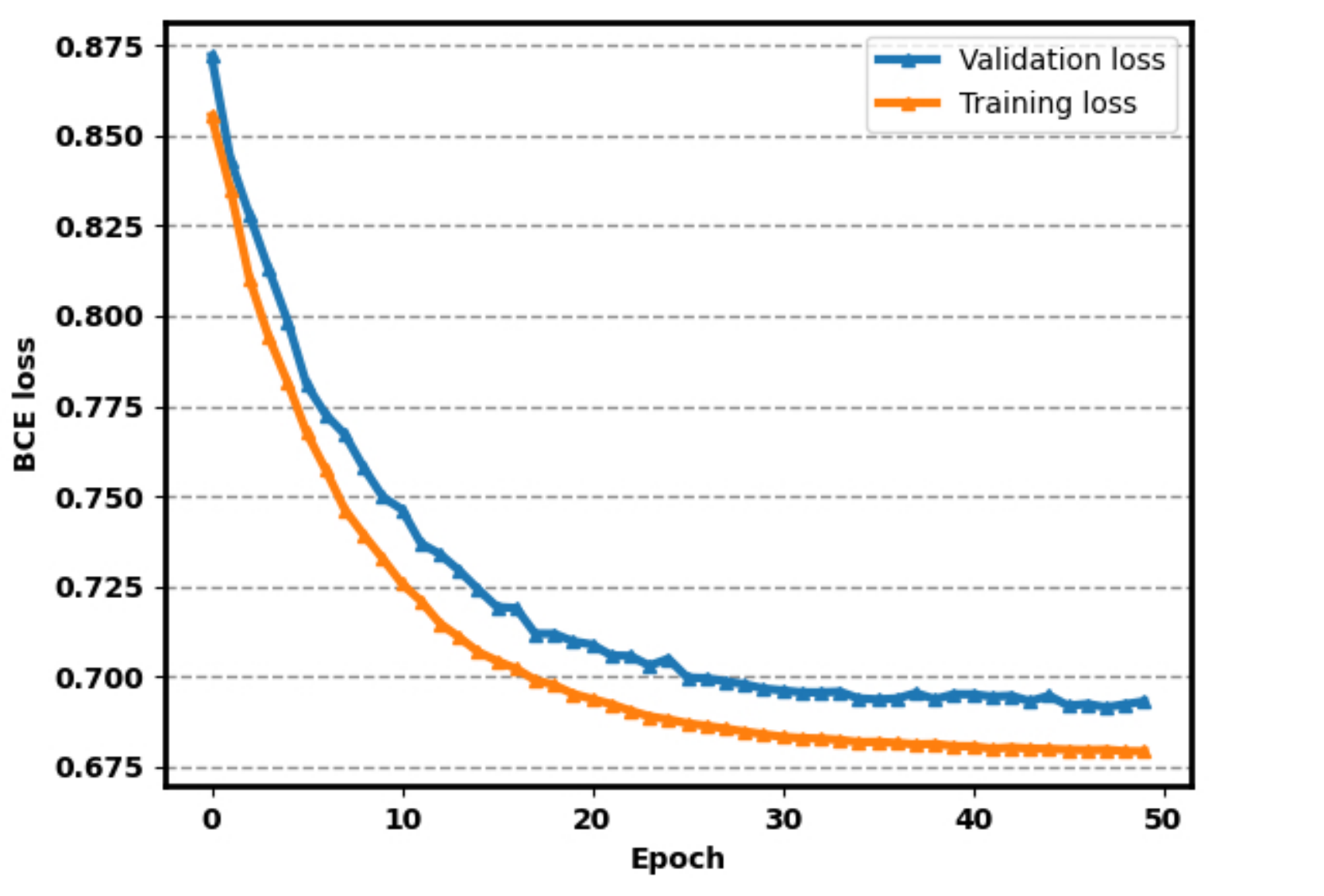}
    \end{minipage}
    }
    \subfigure[Training and validation loss of the semantic loss]{
    \begin{minipage}[b]{0.4\textwidth}
    \includegraphics[width=1\textwidth]{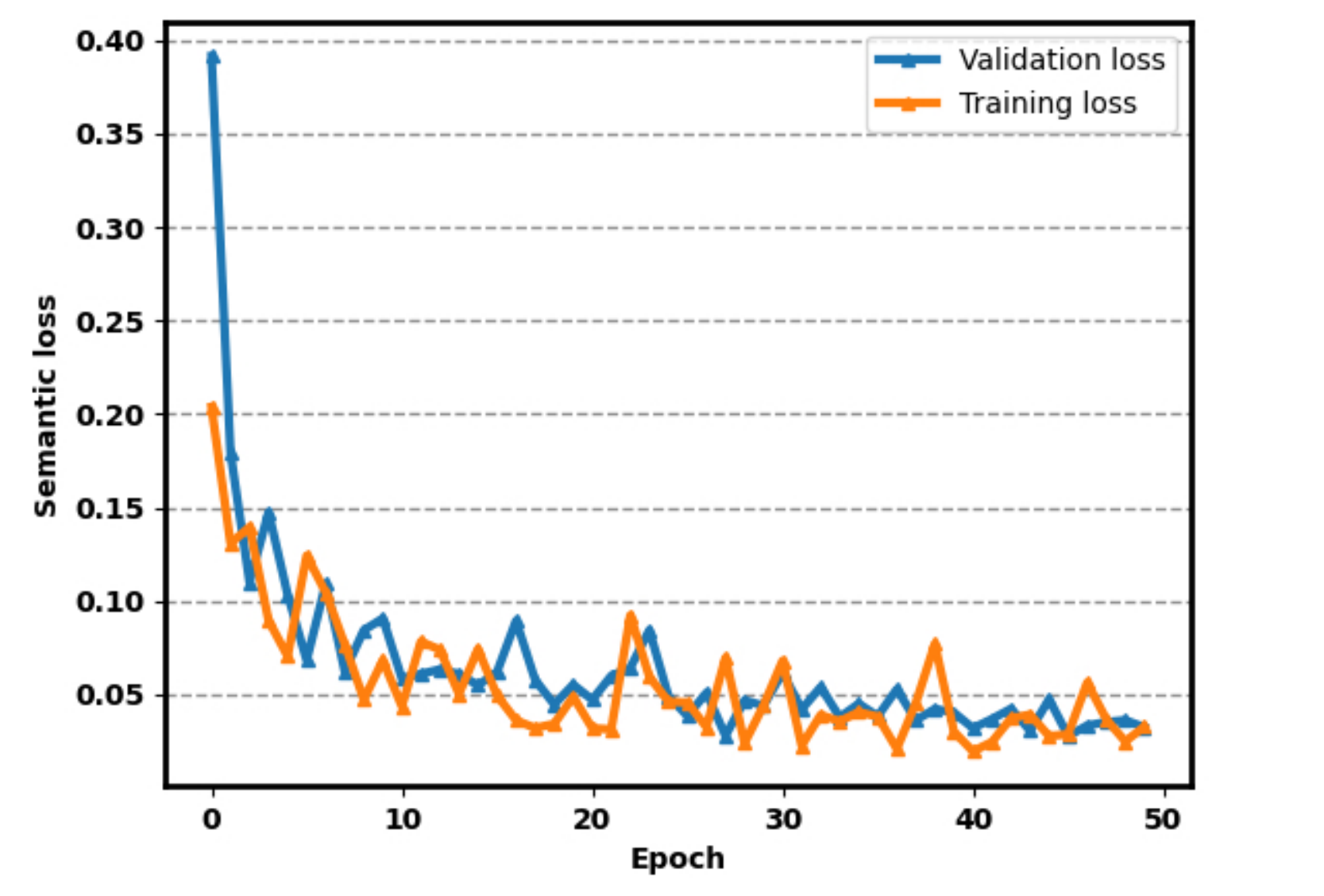}
    \end{minipage}
    }
    \centering
    \caption{BCE loss and semantic loss of the semantic encoder and decoder of table.\ref{mnist_structure} for the MNIST dataset} \label{mnist_loss}
\end{figure}

\begin{figure}
    \centering
    \setlength{\abovecaptionskip}{0.cm}
    
    \subfigure[SSIM with different compression ratio]{
    \begin{minipage}[b]{0.35\textwidth}
    \includegraphics[width=1\textwidth]{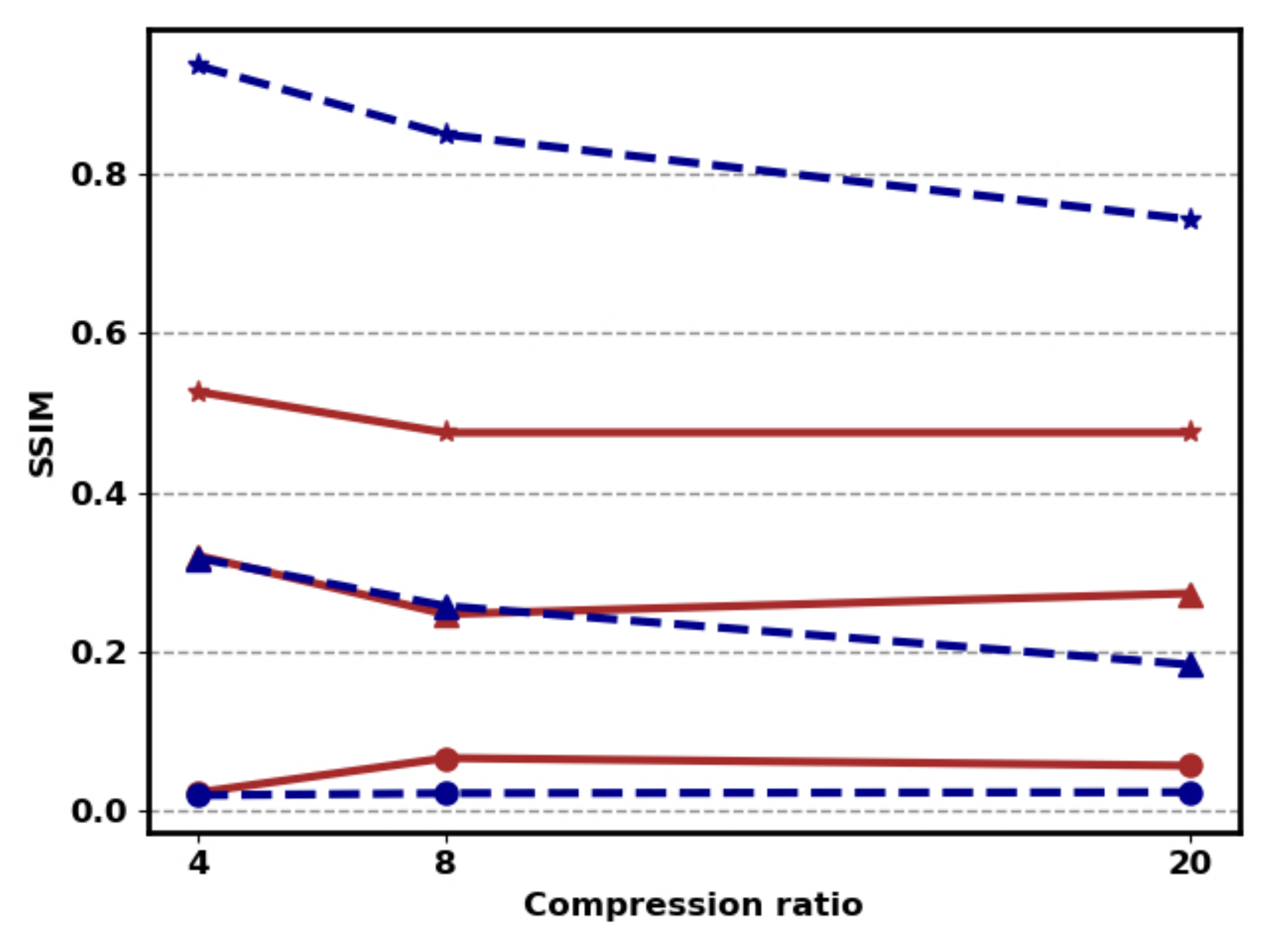}
    \end{minipage}
    }

    \subfigure[PSNR with different compression ratio]{
    \begin{minipage}[b]{0.35\textwidth}
    \includegraphics[width=1\textwidth]{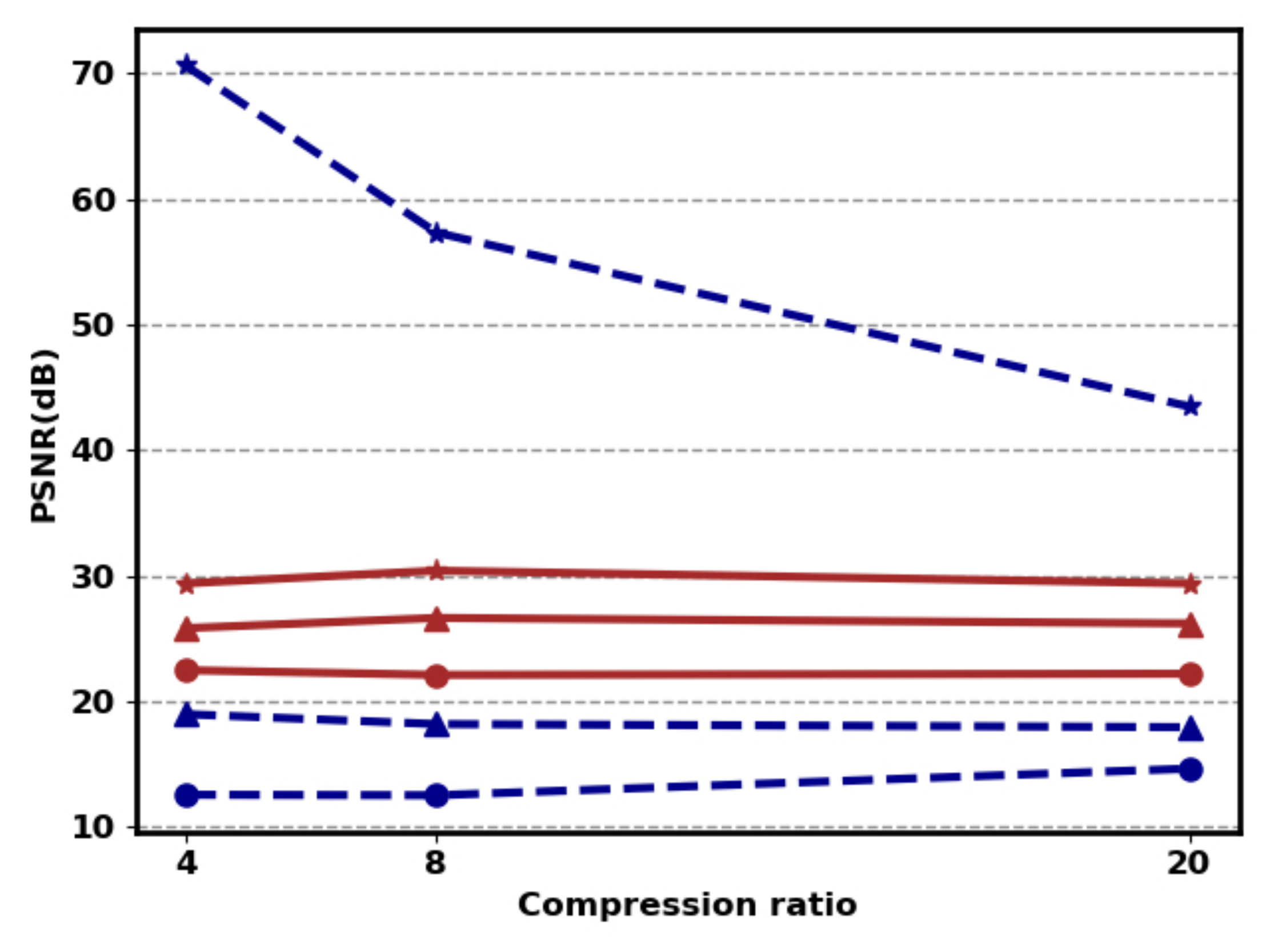}
    \end{minipage}
    }
    
    \subfigure[Recognition rate ratio with different compression ratio]{
    \begin{minipage}[b]{0.35\textwidth}
    \includegraphics[width=1\textwidth]{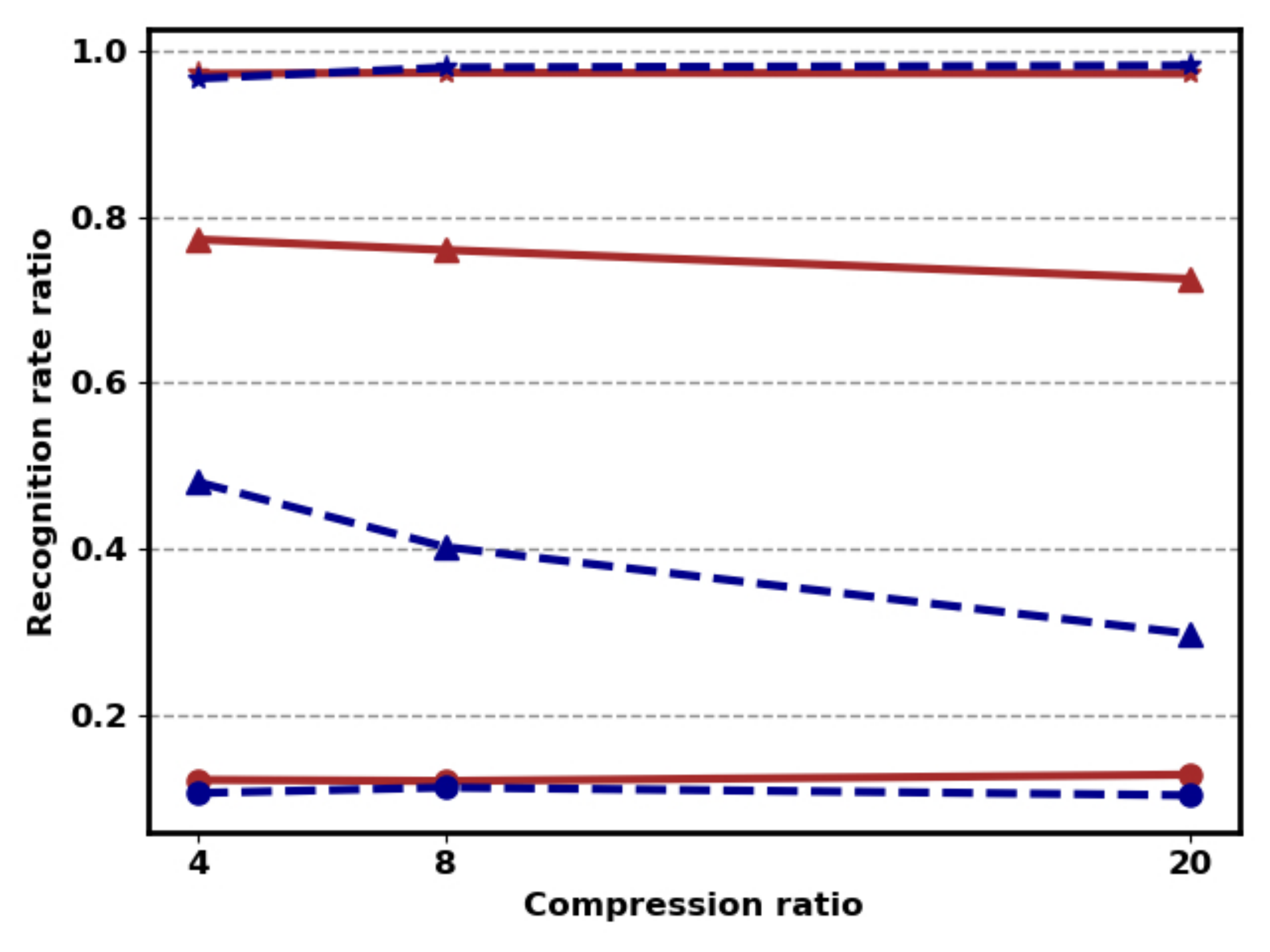}
    \end{minipage}
    }
    
    \subfigure[LPIPS with different compression ratio]{
    \begin{minipage}[b]{0.35\textwidth}
    \includegraphics[width=1\textwidth]{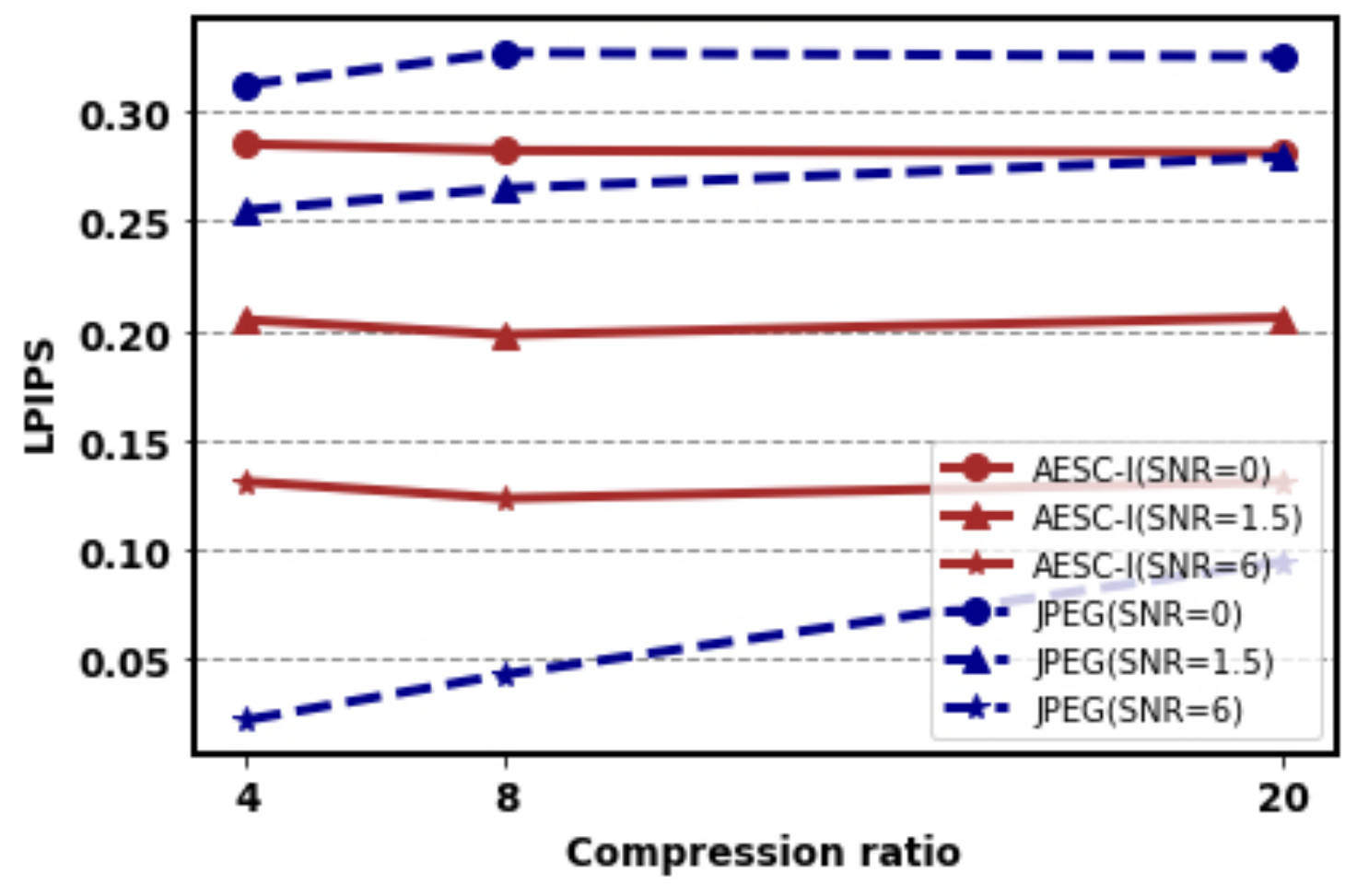}
    \end{minipage}
    }
    
    \caption{Performance comparison between Semantic communication system and JPEG under AWGN channel with different compression rate for the MNIST dataset. The legends of the four subfigures are the same, so they are only presented in (d).} \label{performance_compress}
\end{figure}

\begin{figure}
    \centering
    \setlength{\abovecaptionskip}{0.cm}
    
    \subfigure[LPIPS with different SNR]{
    \begin{minipage}[b]{0.4\textwidth}
    \includegraphics[width=1\textwidth]{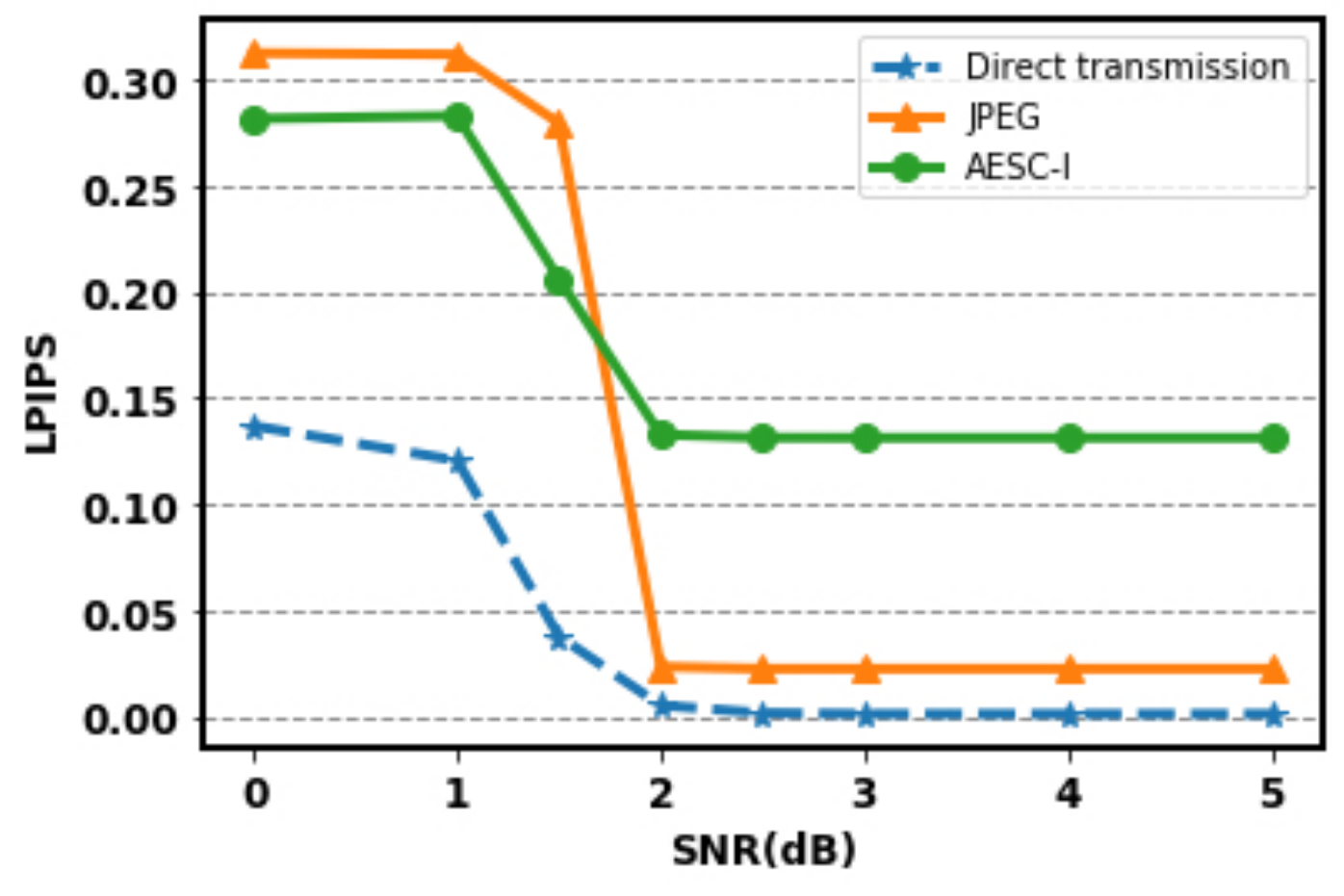}
    \end{minipage}
    }
    
    \subfigure[Recognition rate ratio with different SNR]{
    \begin{minipage}[b]{0.4\textwidth}
    \includegraphics[width=1\textwidth]{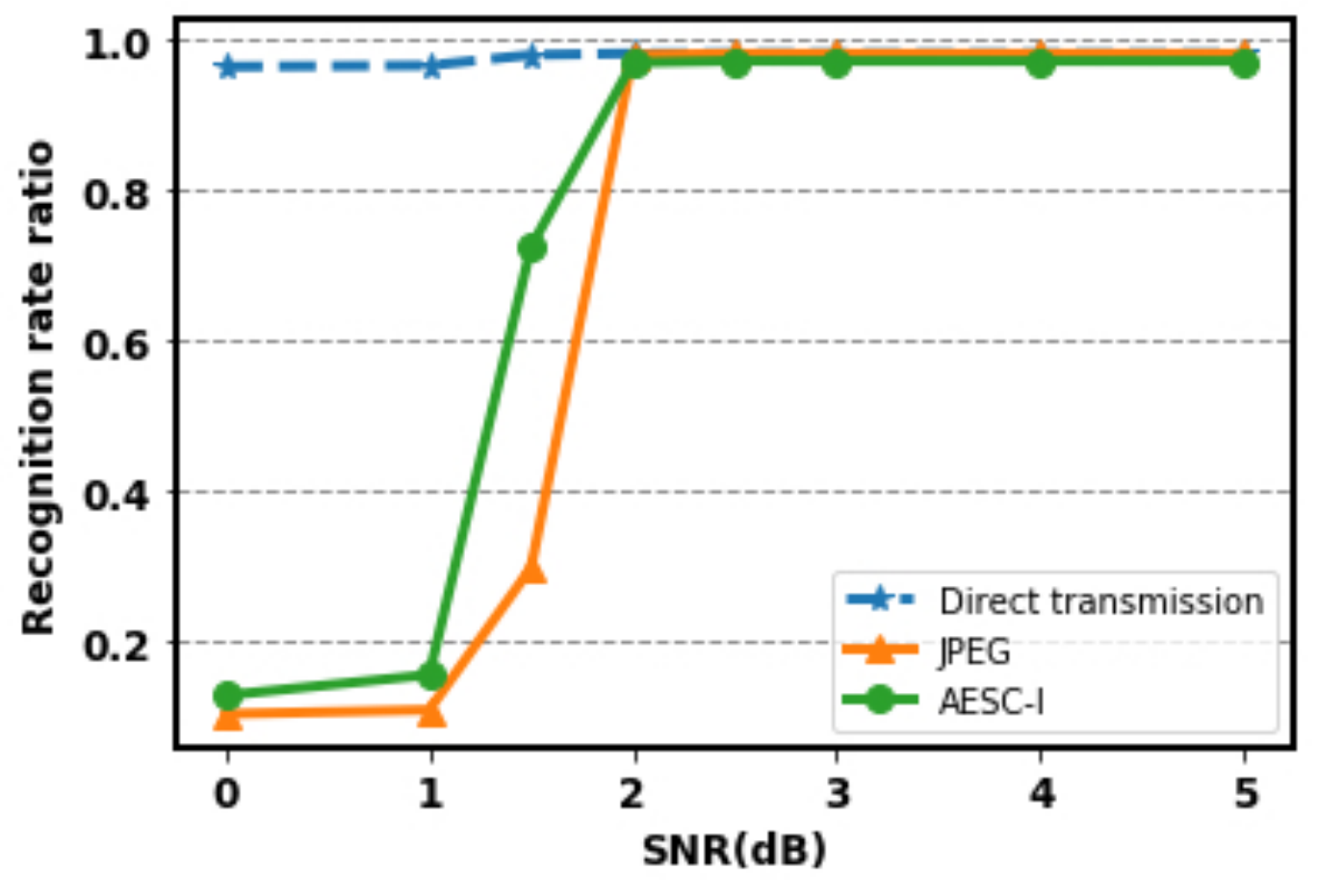}
    \end{minipage}
    }

    \caption{LPIPS and Recognition rate ratio comparison between Semantic communication system and JPEG with 20 times compression rate under AWGN channel with different SNR for the MNIST dataset. Direct transmission means that no compression processing is performed at the source.} \label{performance_snr_mnist}
\end{figure}

\begin{figure}
    \centering
    \setlength{\abovecaptionskip}{0.cm}
    
    \subfigure[LPIPS with different SNR]{
    \begin{minipage}[b]{0.4\textwidth}
    \includegraphics[width=1\textwidth]{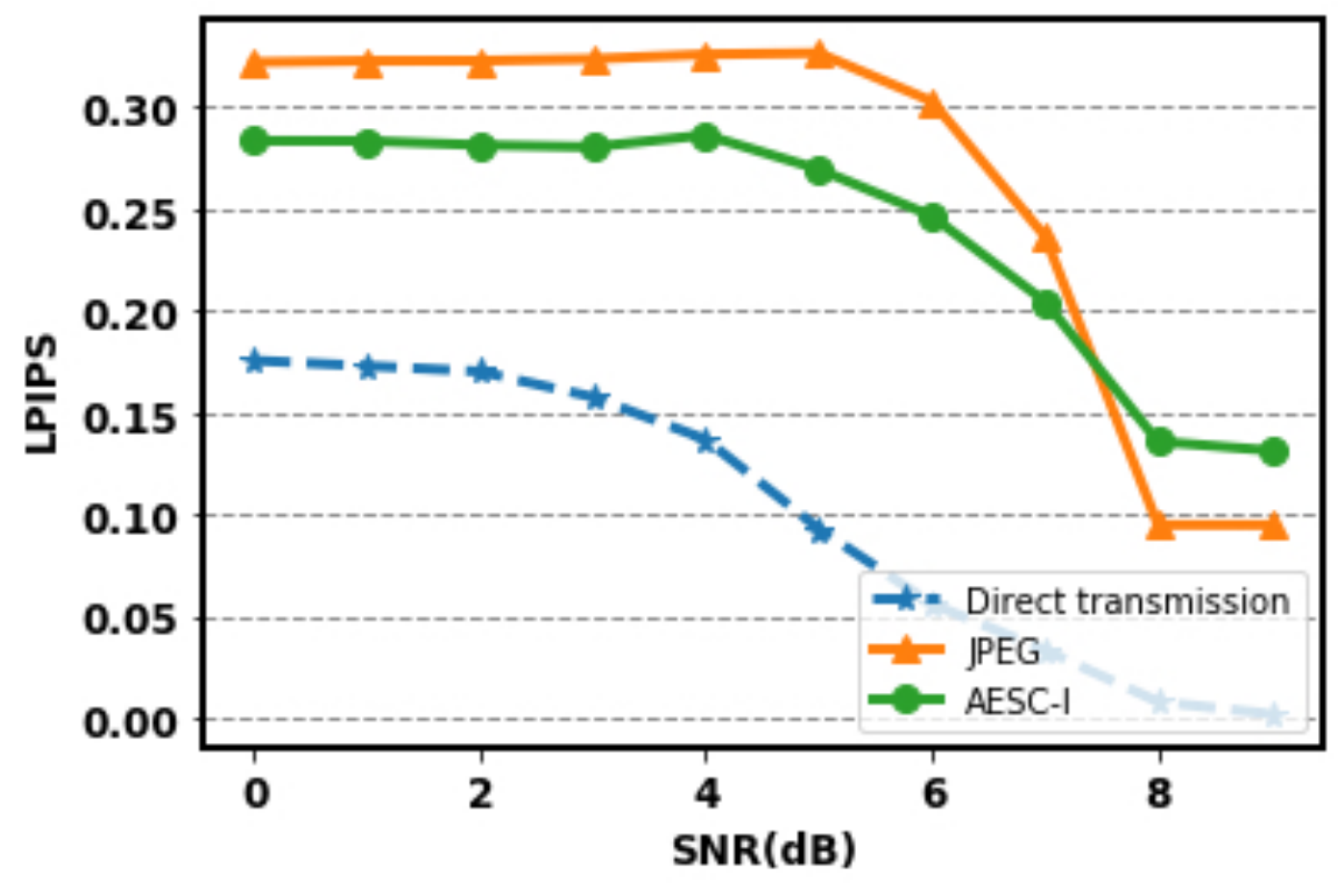}
    \end{minipage}
    }
    
    \subfigure[Recognition rate ratio with different SNR]{
    \begin{minipage}[b]{0.4\textwidth}
    \includegraphics[width=1\textwidth]{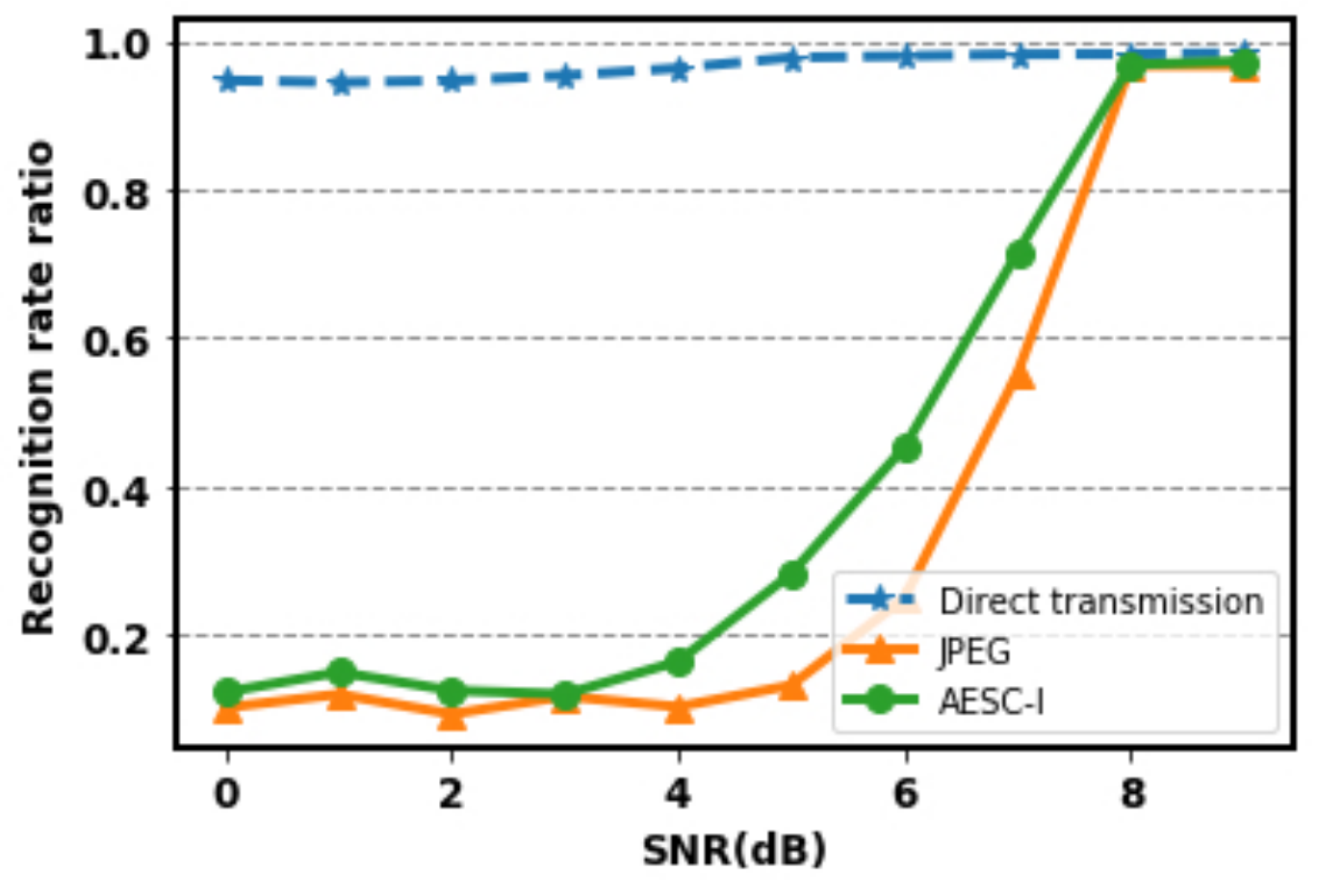}
    \end{minipage}
    }

    \caption{LPIPS and Recognition rate ratio comparison between Semantic communication system and JPEG with 20 times compression rate under the slow fading channel with different SNR for the MNIST dataset. Direct transmission means that no compression processing is performed at the source.} \label{performance_snr_fading_mnist}
\end{figure}

\begin{figure*}
    \centering
    \setlength{\abovecaptionskip}{0.cm}
    
    \subfigure[Reconstructed example of digital "0" under different SNR]{
    \begin{minipage}[b]{1\textwidth}
    \centering
    \includegraphics[width=0.8\textwidth]{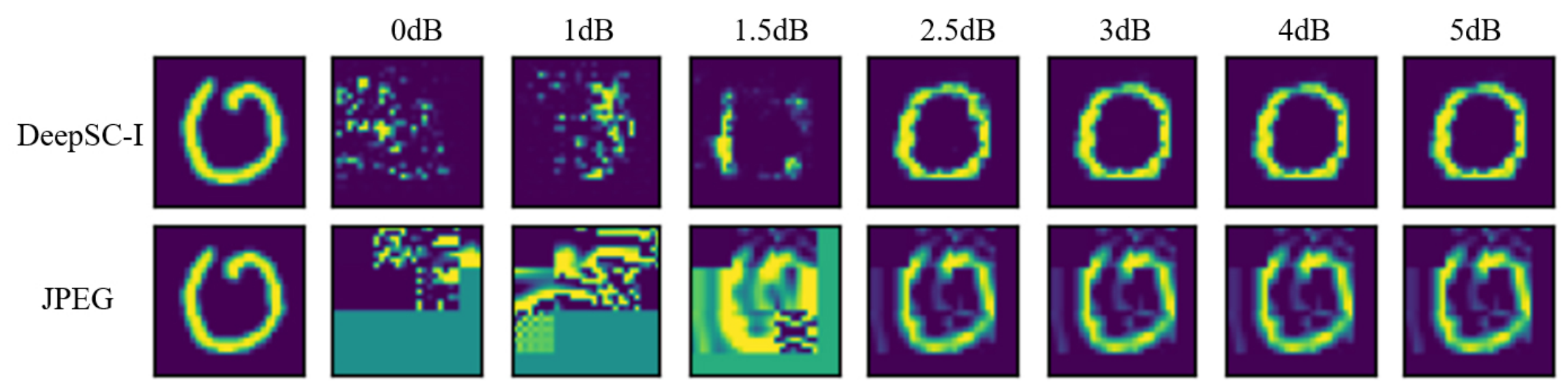}
    \end{minipage}
    }
    
    \subfigure[Reconstructed example of digital "2" under different SNR]{
    \begin{minipage}[b]{1\textwidth}
    \centering
    \includegraphics[width=0.8\textwidth]{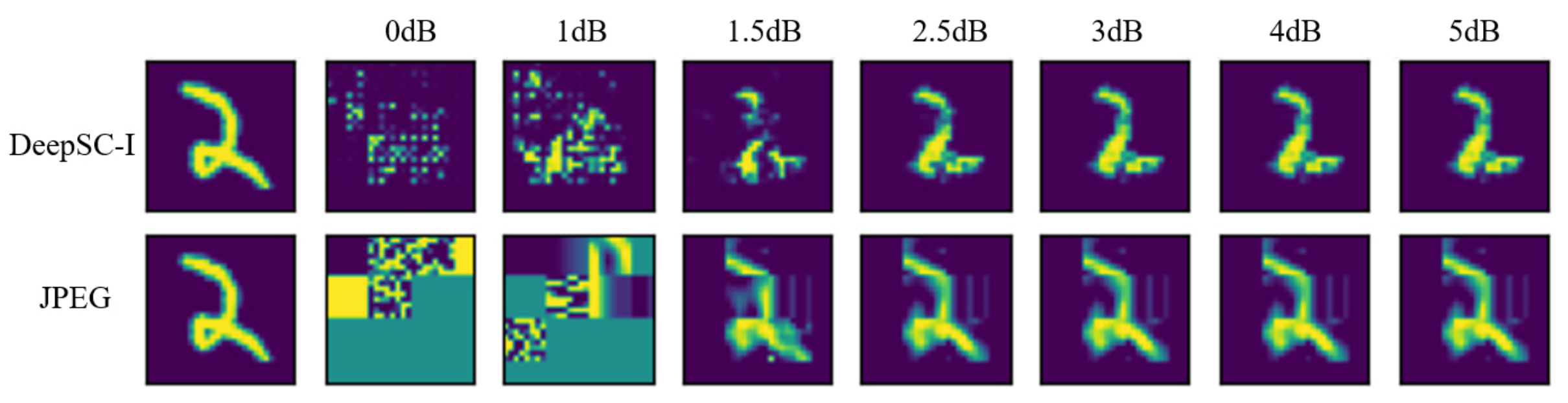}
    \end{minipage}
    }

    \caption{Examples of reconstructed images produced by the AESC-I and JPEG under AWGN channel with the 20 times compression ratio.} \label{mnist_example_awgn}
\end{figure*}

\begin{figure*}
    \centering
    \setlength{\abovecaptionskip}{0.cm}
    
    \subfigure[Reconstructed example of digital "6" under different SNR]{
    \begin{minipage}[b]{1\textwidth}
    \centering
    \includegraphics[width=0.8\textwidth]{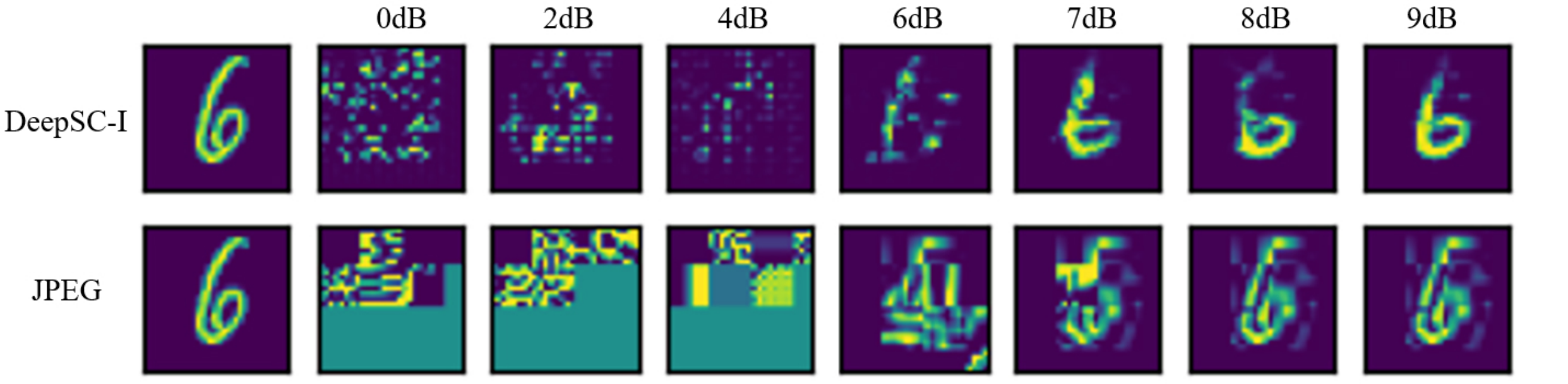}
    \end{minipage}
    }
    
    \subfigure[Reconstructed example of digital "8" under different SNR]{
    \begin{minipage}[b]{1\textwidth}
    \centering
    \includegraphics[width=0.8\textwidth]{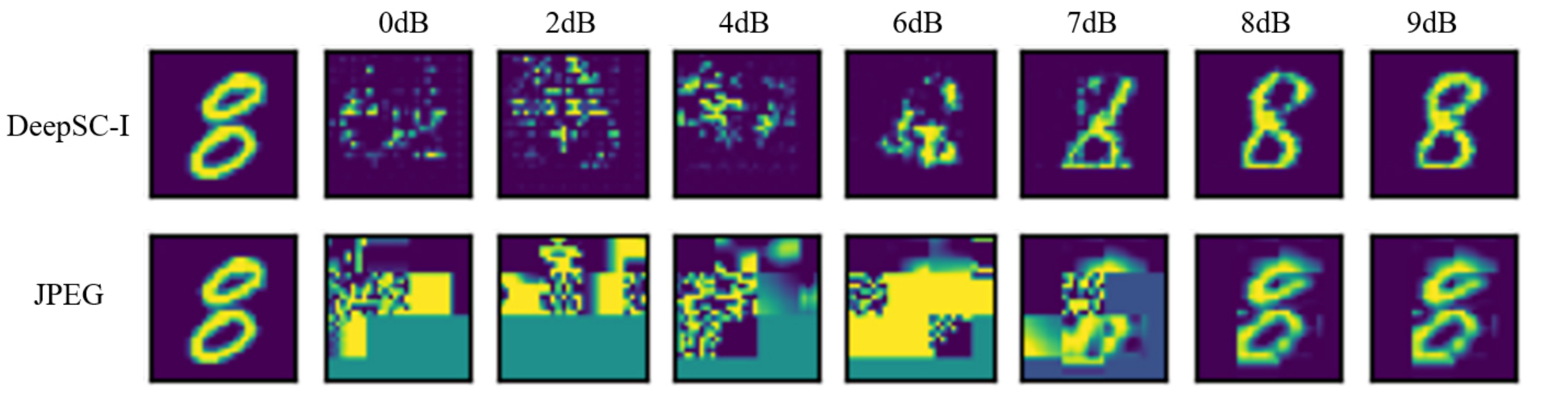}
    \end{minipage}
    }

    \caption{Examples of reconstructed images produced by the AESC-I and JPEG under the slow fading channel with the 20 times compression ratio.} \label{mnist_example_fading}
\end{figure*}

\begin{figure}
    \centering
    \setlength{\abovecaptionskip}{0.cm}
    \subfigure[Training and validation loss of the MSE loss]{
    \begin{minipage}[b]{0.4\textwidth}
    \includegraphics[width=1\textwidth]{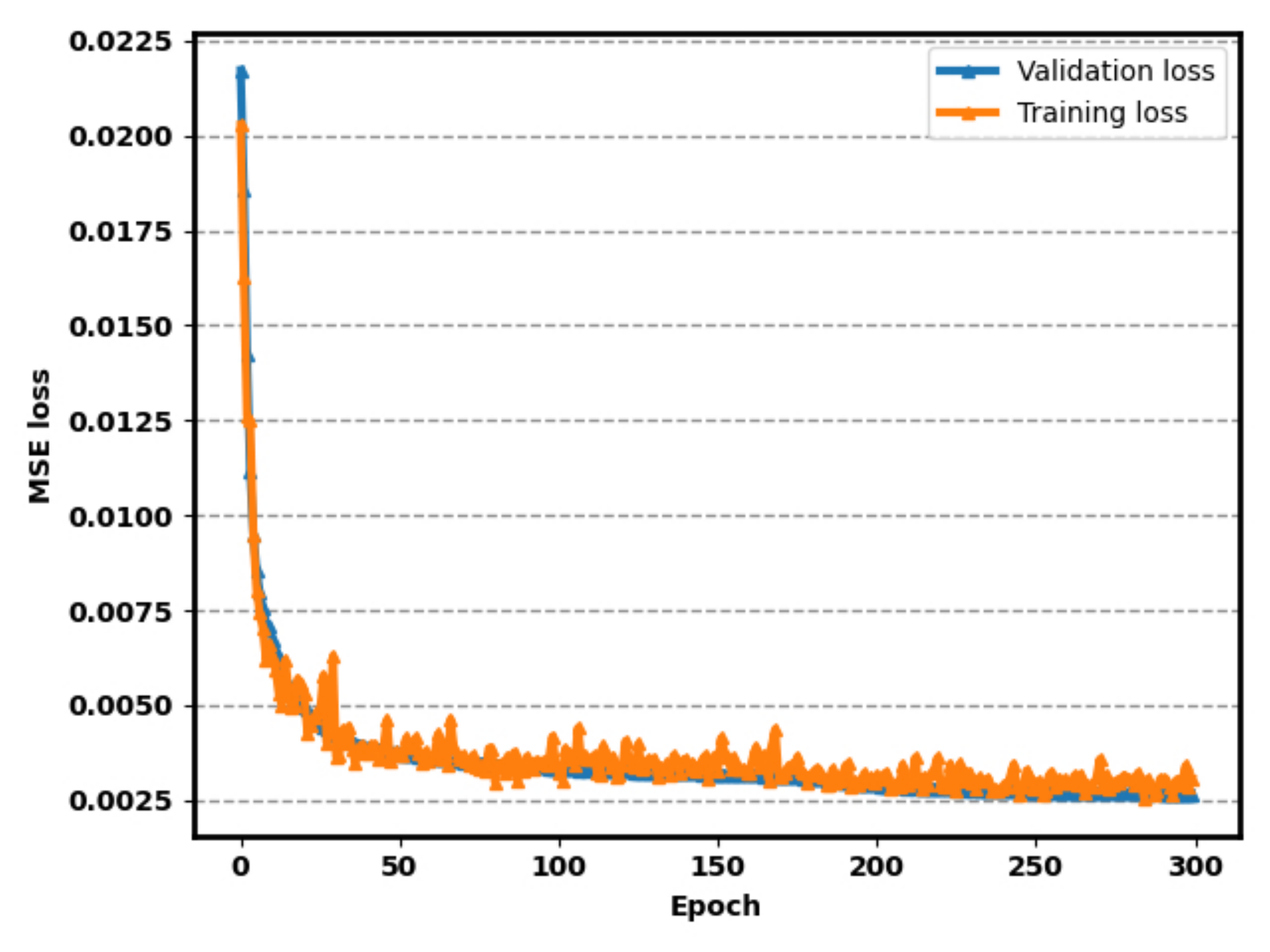}
    \end{minipage}
    }
    \subfigure[Training and validation loss of the semantic loss]{
    \begin{minipage}[b]{0.4\textwidth}
    \includegraphics[width=1\textwidth]{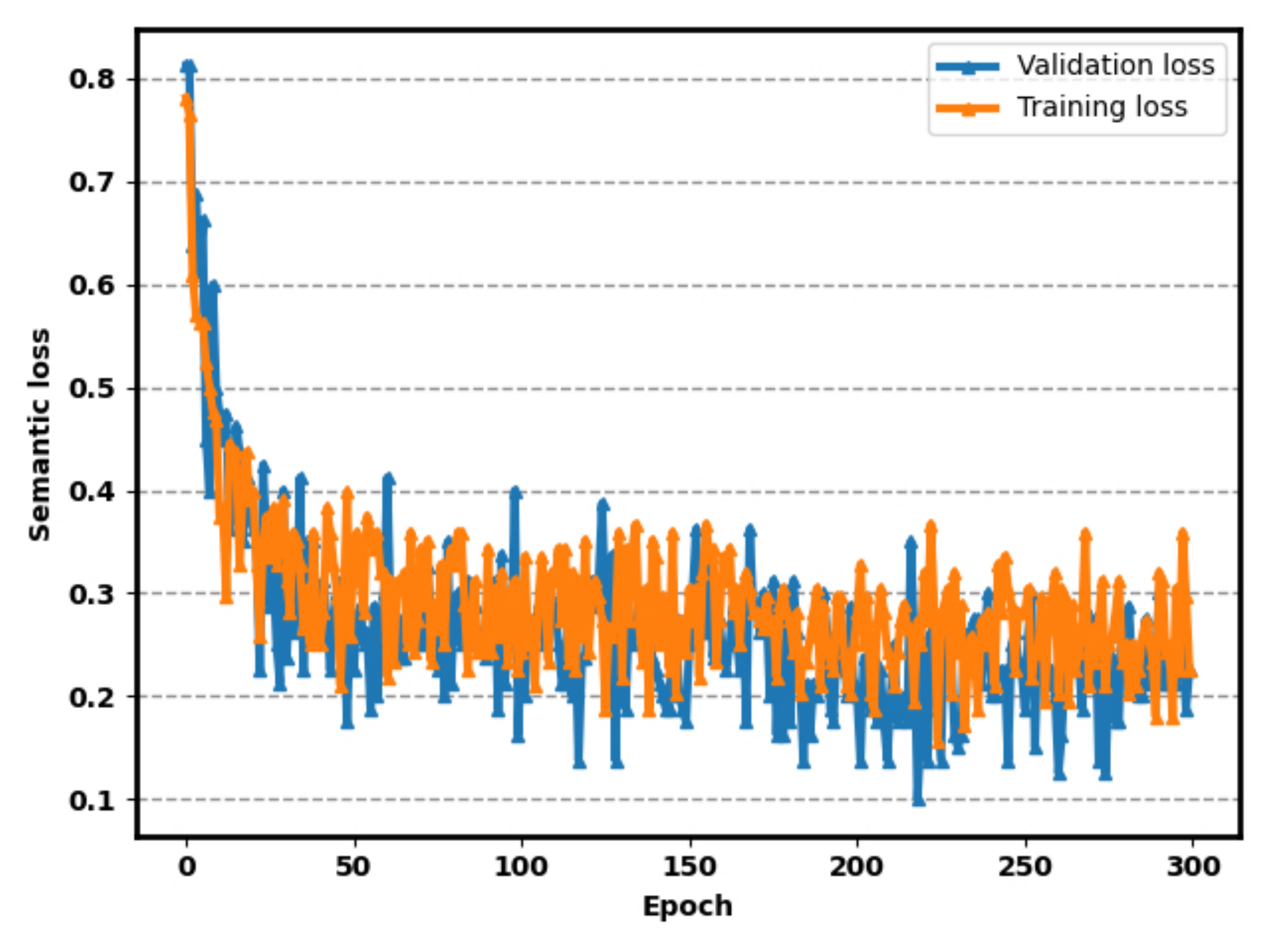}
    \end{minipage}
    }
    \centering
    \caption{MSE loss and semantic loss of the semantic encoder and decoder of table.\ref{cifar_structure} for the Cifar-10 dataset} \label{cifar_loss}
\end{figure}

\begin{figure}
    \centering
    \setlength{\abovecaptionskip}{0.cm}
    \subfigure[Recognition rate ratio with different compression ratio]{
    \begin{minipage}[b]{0.4\textwidth}
    \includegraphics[width=1\textwidth]{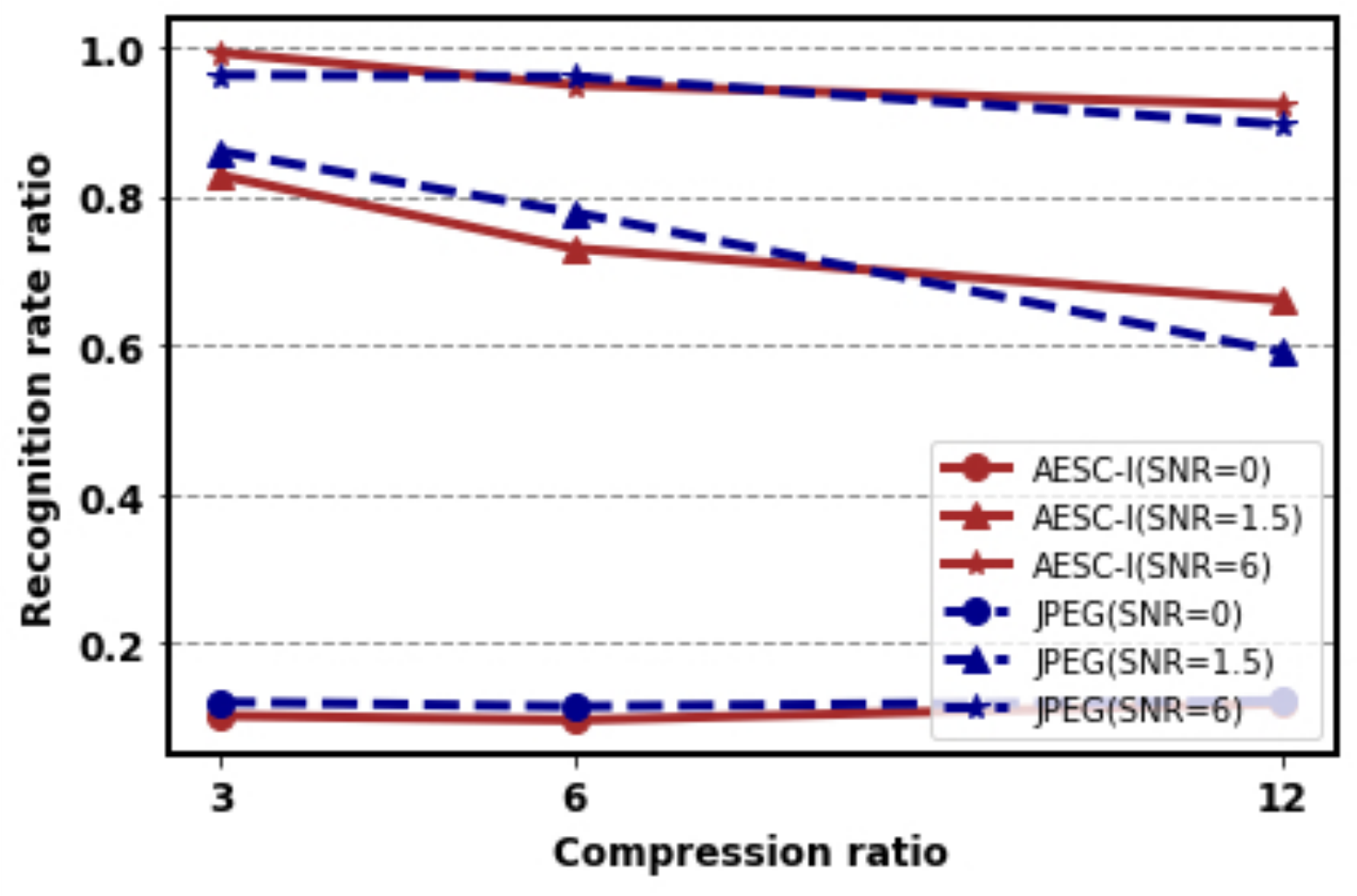}
    \end{minipage}
    }
    \subfigure[LPIPS with different compression ratio]{
    \begin{minipage}[b]{0.4\textwidth}
    \includegraphics[width=1\textwidth]{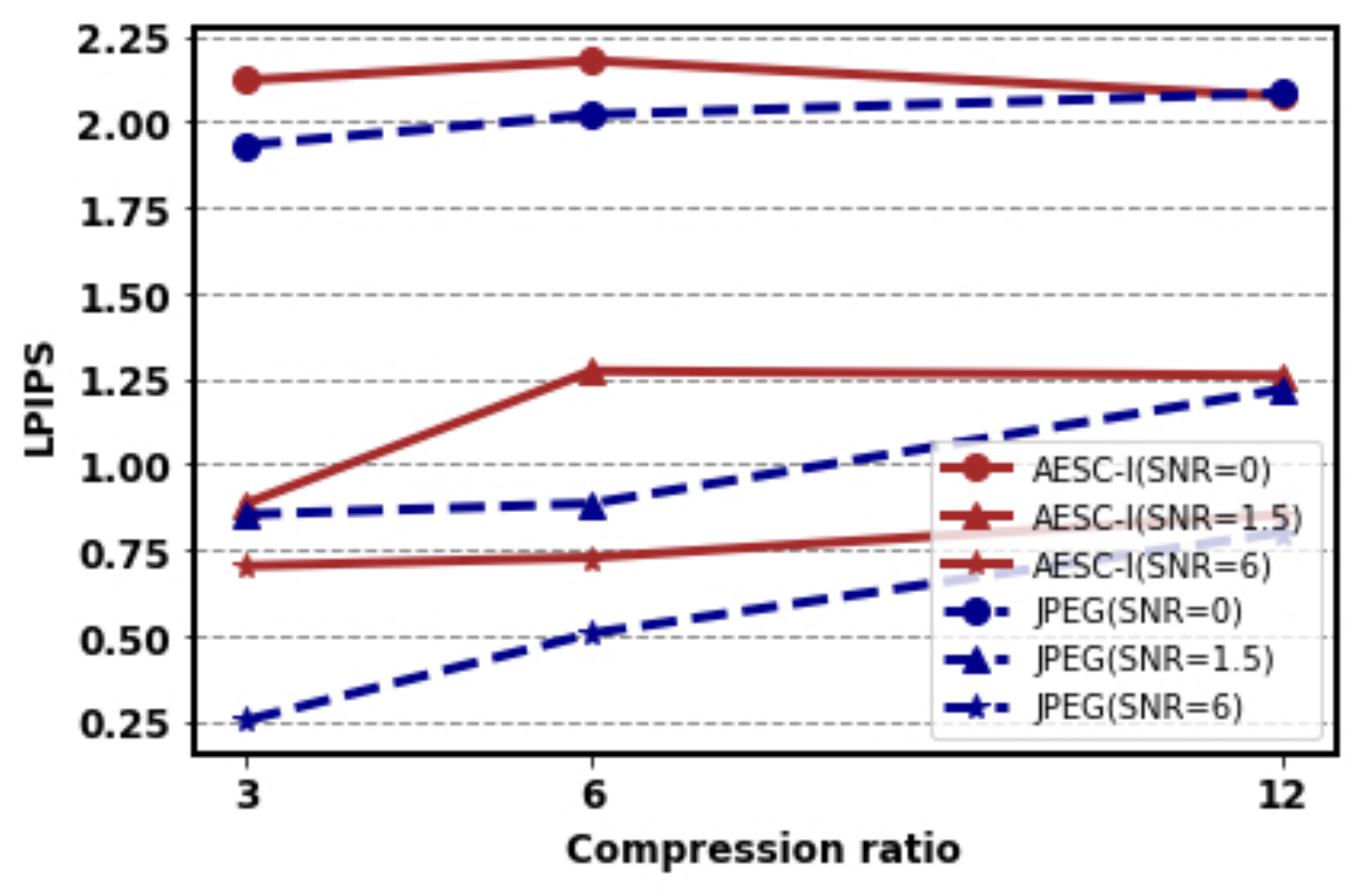}
    \end{minipage}
    }
    \centering
    \caption{Performance comparison between Semantic communication system and JPEG under AWGN channel with different compression rate for the Cifar-10 dataset. } \label{performance_compress_cifar}
\end{figure}

\begin{figure}
    \centering
    \setlength{\abovecaptionskip}{0.cm}
    \subfigure[Recognition rate ratio with different SNR]{
    \begin{minipage}[b]{0.4\textwidth}
    \includegraphics[width=1\textwidth]{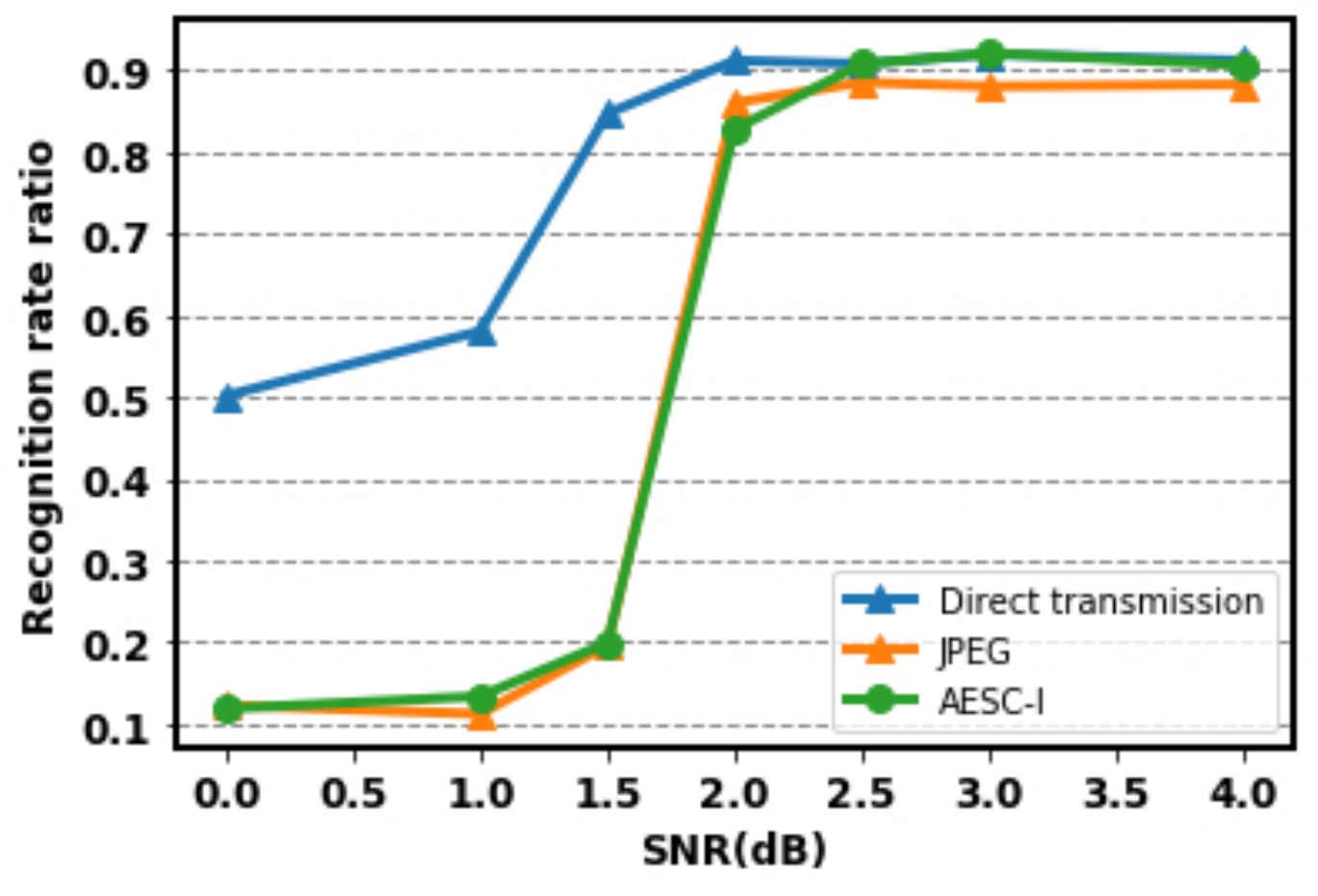}
    \end{minipage}
    }
    \subfigure[LPIPS with different SNR]{
    \begin{minipage}[b]{0.4\textwidth}
    \includegraphics[width=1\textwidth]{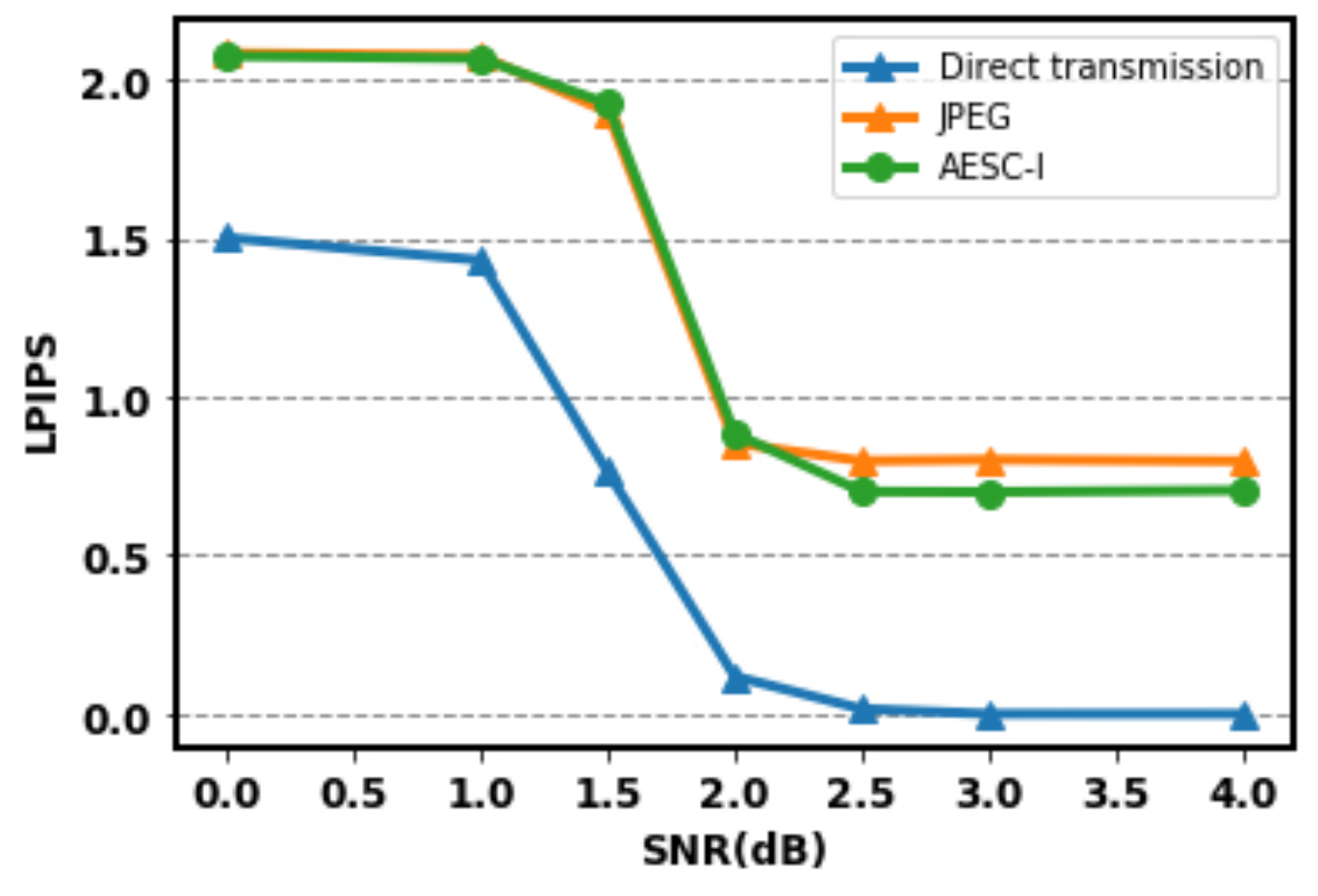}
    \end{minipage}
    }
    \centering
    \caption{ Recognition rate ratio and LPIPS comparison between Semantic communication system and JPEG with 12 times compression rate under AWGN channel with different SNR for the Cifar-10.} \label{snr_performance_awgn_cifar}
\end{figure}

\begin{figure}
    \centering
    \setlength{\abovecaptionskip}{0.cm}
    \subfigure[Recognition rate ratio with different SNR]{
    \begin{minipage}[b]{0.4\textwidth}
    \includegraphics[width=1\textwidth]{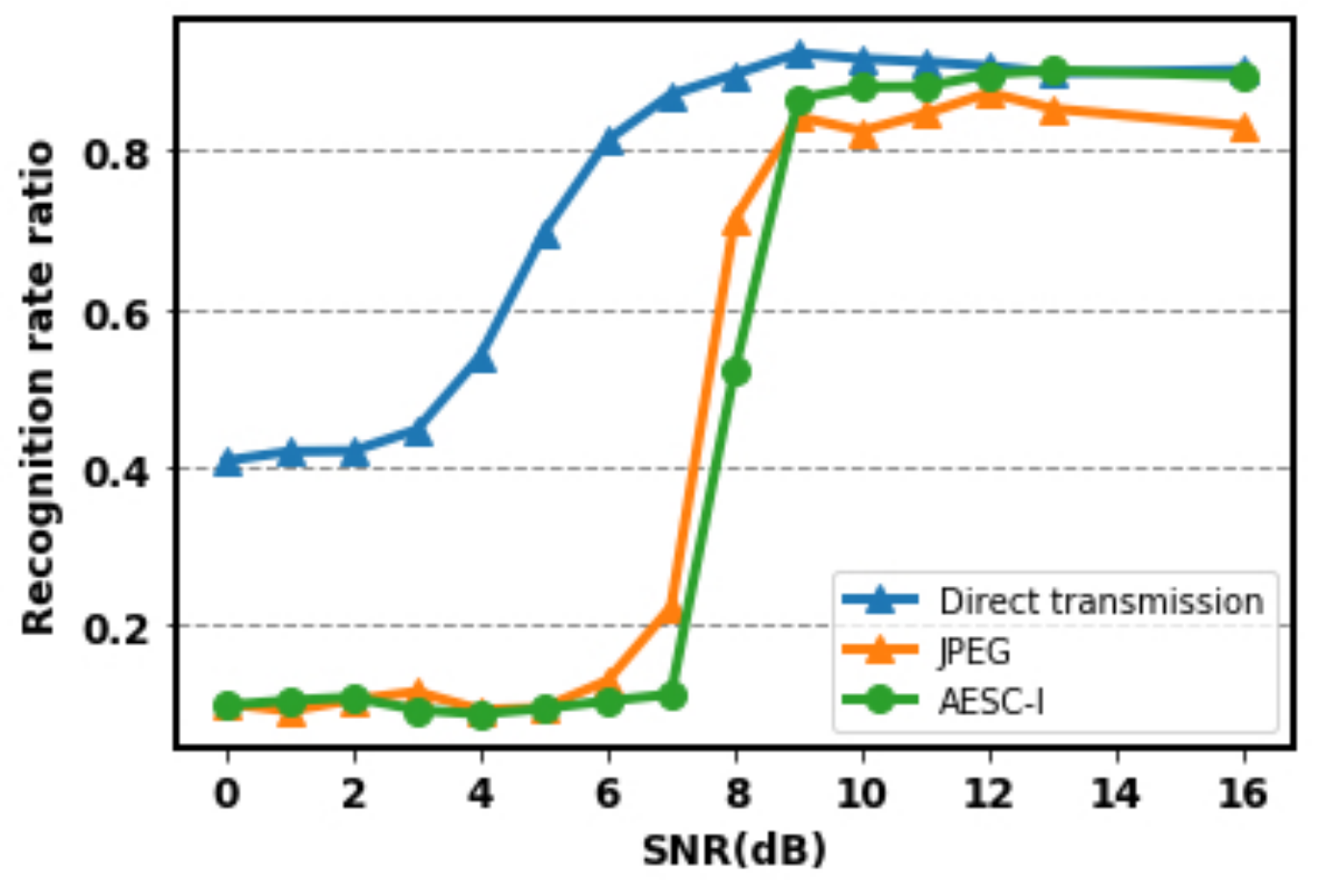}
    \end{minipage}
    }
    \subfigure[LPIPS with different SNR]{
    \begin{minipage}[b]{0.4\textwidth}
    \includegraphics[width=1\textwidth]{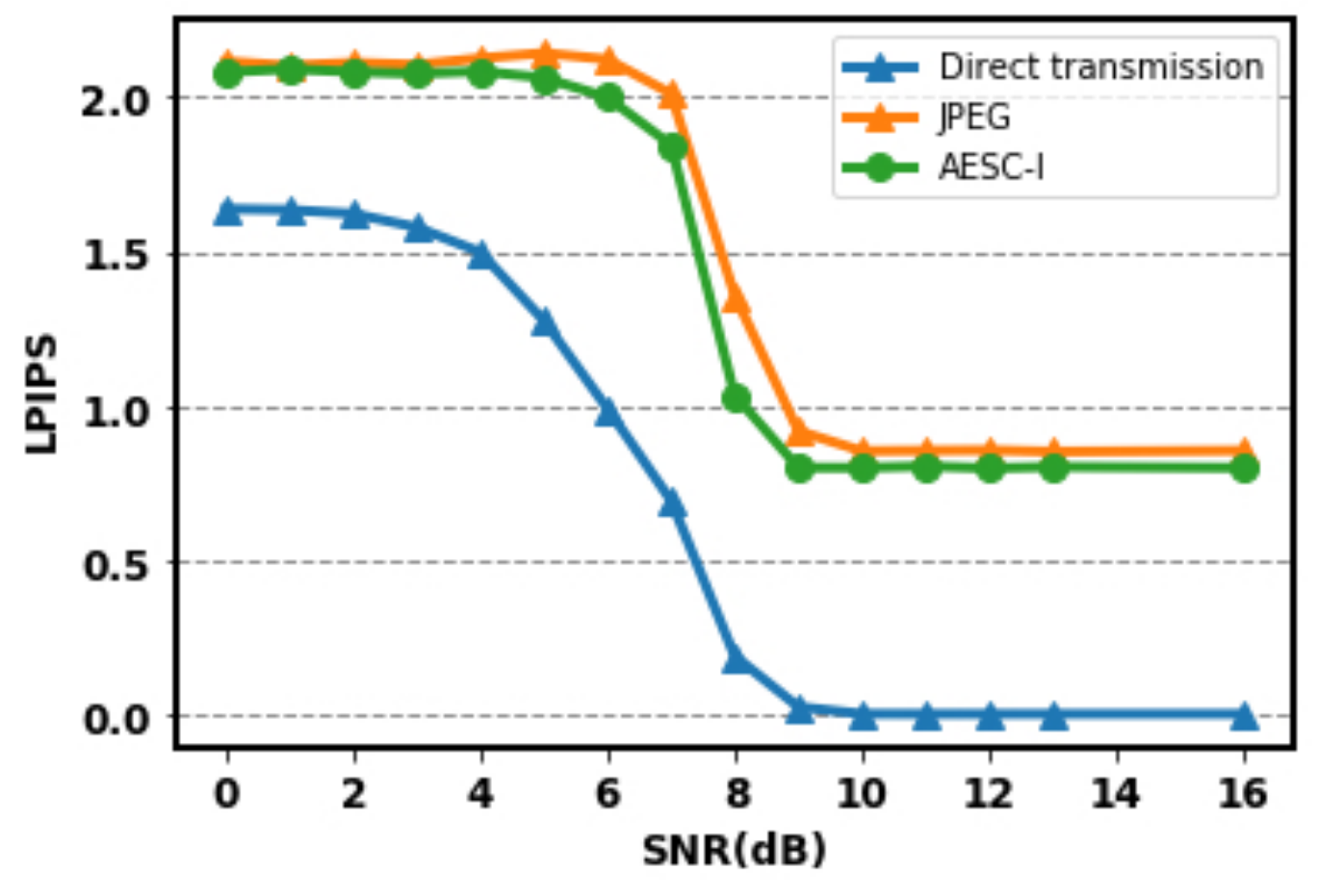}
    \end{minipage}
    }
    \centering
    \caption{ Recognition rate ratio and LPIPS comparison between Semantic communication system and JPEG with 12 times compression rate under the show fading channel with different SNR for the Cifar-10.} \label{snr_performance_fading_cifar}
\end{figure}

\begin{figure*}
    \centering
    \setlength{\abovecaptionskip}{0.cm}
    
    \subfigure[Reconstructed example of "horse" under different SNR]{
    \begin{minipage}[b]{1\textwidth}
    \includegraphics[width=1\textwidth]{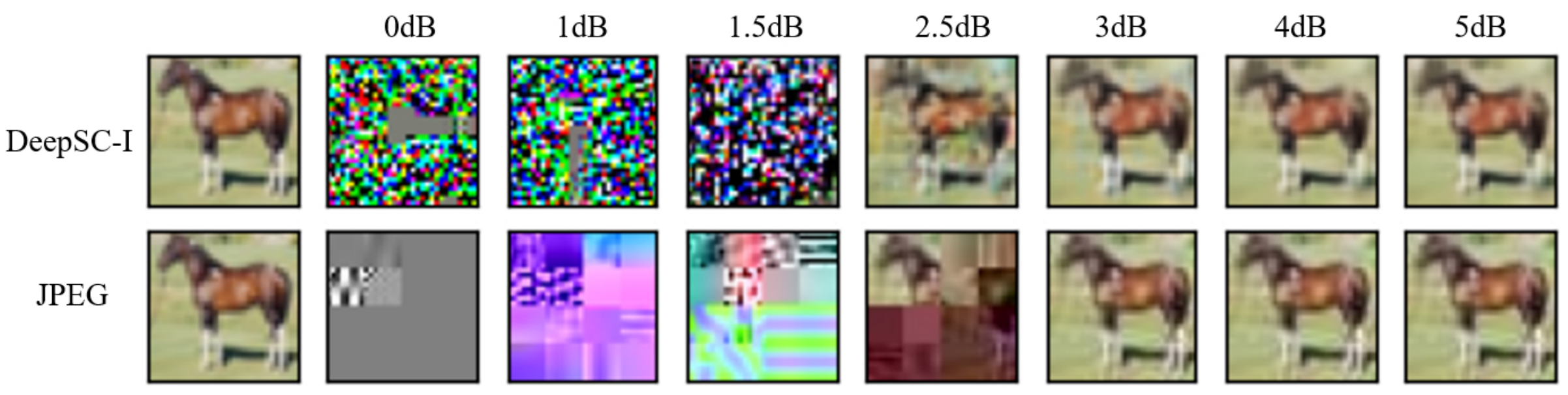}
    \end{minipage}
    }
    
    \subfigure[Reconstructed example of "car" under different SNR]{
    \begin{minipage}[b]{1\textwidth}
    \includegraphics[width=1\textwidth]{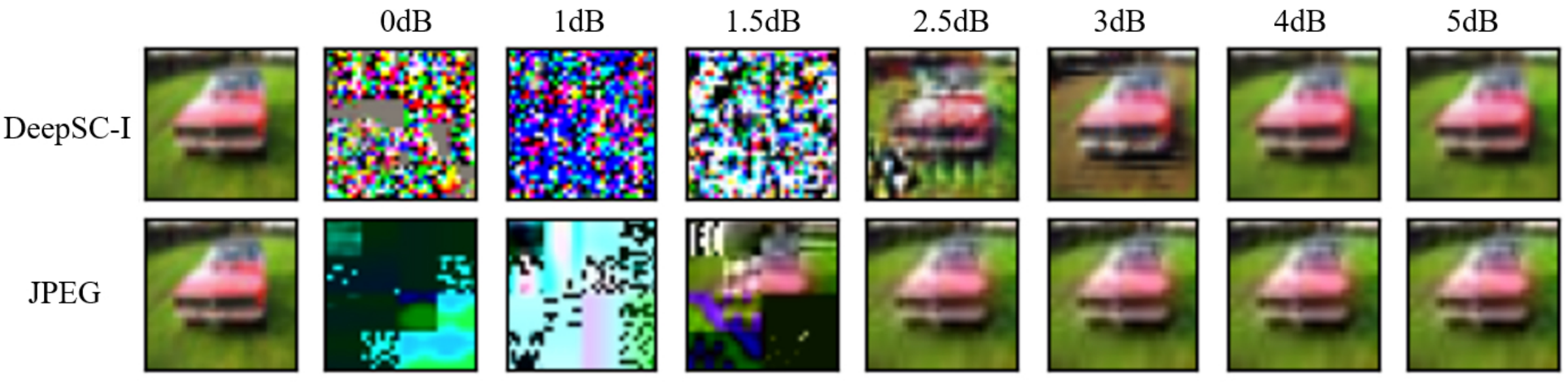}
    \end{minipage}
    }

    \caption{Examples of reconstructed images produced by the AESC-I and JPEG under the AWGN channel with the 12 times compression ratio.} \label{cifar_example_awgn}
\end{figure*}

\begin{figure*}
    \centering
    \setlength{\abovecaptionskip}{0.cm}
    
    \subfigure[Reconstructed example of "horse" under different SNR]{
    \begin{minipage}[b]{1\textwidth}
    \includegraphics[width=1\textwidth]{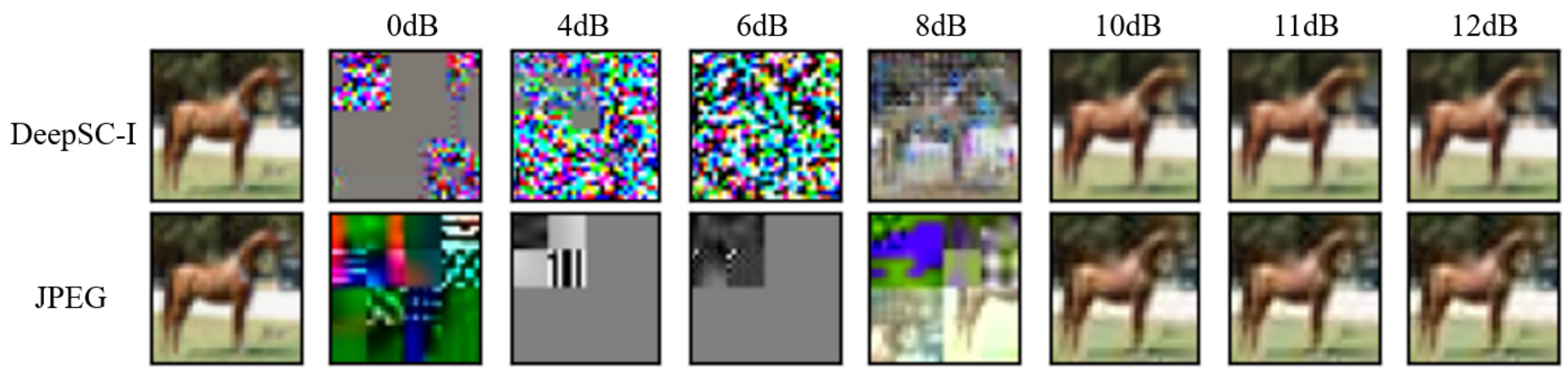}
    \end{minipage}
    }
    
    \subfigure[Reconstructed example of "bus" under different SNR]{
    \begin{minipage}[b]{1\textwidth}
    \includegraphics[width=1\textwidth]{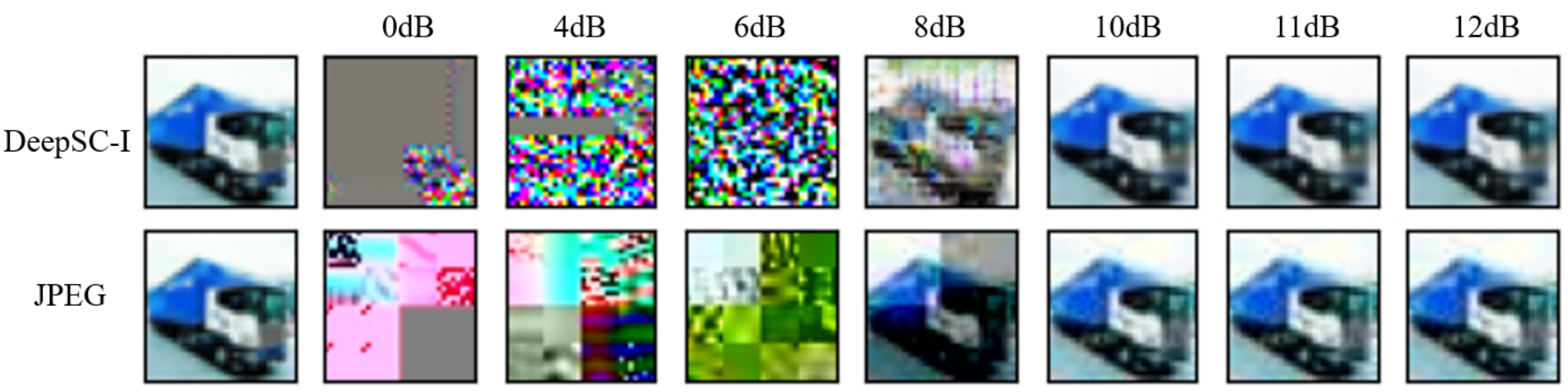}
    \end{minipage}
    }

    \caption{Examples of reconstructed images produced by the AESC-I and JPEG under the slow fading channel with the 12 times compression ratio.} \label{cifar_example_fading}
\end{figure*}
\begin{table}[]
\setlength{\abovecaptionskip}{0.cm}
\centering
\caption{Proposed semantic encoder and decoder for MNIST}\label{mnist_structure}
\begin{tabular}{@{}|c|c|@{}}
\toprule
\multicolumn{1}{|l|}{} & Structure \\ \midrule
\multirow{4}{*}{Encoder} & \begin{tabular}[c]{@{}c@{}}Conv2d,   Kernel size=(3, 3), stride=(2, 2)\\    \\ input   channels=1, output channels=16, ReLU\end{tabular} \\ \cmidrule(l){2-2} 
 & \begin{tabular}[c]{@{}c@{}}Conv2d,Kernel   size=(3, 3), stride=(2, 2)\\    \\ input   channels=16, output channels=16, ReLU\end{tabular} \\ \cmidrule(l){2-2} 
 & \begin{tabular}[c]{@{}c@{}}Conv2d,Kernel   size=(3, 3), stride=(2, 2)\\    \\ input   channels=16, output channels=16, ReLU\end{tabular} \\ \cmidrule(l){2-2} 
 & Linear,   input features=16, output features=$Z_{dims}$, ReLU \\ \midrule
\multirow{5}{*}{Decoder} & Linear,   input features=$Z_{dims}$, output features=64, ReLU \\ \cmidrule(l){2-2} 
 & \begin{tabular}[c]{@{}c@{}}ConvTran2D,   kernel size=(3, 3), stride=(2, 2),\\    \\ input   channels=64, output channels=16, ReLU\end{tabular} \\ \cmidrule(l){2-2} 
 & \begin{tabular}[c]{@{}c@{}}ConvTran2D,   kernel size=(3, 3), stride=(2, 2),\\    \\ input   channels=16, output channels=16, ReLU\end{tabular} \\ \cmidrule(l){2-2} 
 & \begin{tabular}[c]{@{}c@{}}ConvTran2D,   kernel size=(3, 3), stride=(2, 2),\\    \\ input   channels=16, output channels=1, Sigmoid\end{tabular} \\ \bottomrule
\end{tabular}
\end{table}

\begin{table}[]
\setlength{\abovecaptionskip}{0.cm}
\centering
\caption{Proposed semantic encoder and decoder for Cifar-10}\label{cifar_structure}
\begin{tabular}{@{}|c|c|@{}}
\toprule
 & Structure \\ \midrule
\multirow{8}{*}{Encoder} & \begin{tabular}[c]{@{}c@{}}Conv2d,   Kernel size=(4, 4), stride=(2, 2)\\    \\ input   channels=3, output channels=32, ReLU\end{tabular} \\ \cmidrule(l){2-2} 
 & \begin{tabular}[c]{@{}c@{}}Conv2d,   Kernel size=(3, 3), stride=(1, 1)\\    \\ input   channels=32, output channels=64, ReLU\end{tabular} \\ \cmidrule(l){2-2} 
 & \begin{tabular}[c]{@{}c@{}}Conv2d,   Kernel size=(3, 3), stride=(1, 1)\\    \\ input   channels=64, output channels=128, ReLU\end{tabular} \\ \cmidrule(l){2-2} 
 & \begin{tabular}[c]{@{}c@{}}Conv2d,   Kernel size=(3, 3), stride=(1, 1)\\    \\ input   channels=128, output channels=256, ReLU\end{tabular} \\ \cmidrule(l){2-2} 
 & \begin{tabular}[c]{@{}c@{}}Conv2d,   Kernel size=(4, 4), stride=(2, 2)\\    \\ input   channels=256, output channels=128, ReLU\end{tabular} \\ \cmidrule(l){2-2} 
 & \begin{tabular}[c]{@{}c@{}}Conv2d,   Kernel size=(3, 3), stride=(1, 1)\\    \\ input   channels=128, output channels=64, ReLU\end{tabular} \\ \cmidrule(l){2-2} 
 & \begin{tabular}[c]{@{}c@{}}Conv2d,   Kernel size=(3, 3), stride=(1, 1)\\    \\ input   channels=64, output channels=32, ReLU\end{tabular} \\ \cmidrule(l){2-2} 
 & \begin{tabular}[c]{@{}c@{}}Conv2d,   Kernel size=(3, 3), stride=(1, 1)\\    \\ input   channels=32, output channels=$Z_{dims}$, ReLU\end{tabular} \\ \midrule
\multirow{7}{*}{Decoder} & \begin{tabular}[c]{@{}c@{}}ConvTran2D,   kernel size=(4, 4), stride=(2, 2),\\    \\ input   channels=$Z_{dims}$, output channels=64, ReLU\end{tabular} \\ \cmidrule(l){2-2} 
 & \begin{tabular}[c]{@{}c@{}}Conv2d,   Kernel size=(3, 3), stride=(1, 1)\\    \\ input   channels=64, output channels=128, ReLU\end{tabular} \\ \cmidrule(l){2-2} 
 & \begin{tabular}[c]{@{}c@{}}Conv2d,   Kernel size=(3, 3), stride=(1, 1)\\    \\ input   channels=128, output channels=128, ReLU\end{tabular} \\ \cmidrule(l){2-2} 
 & \begin{tabular}[c]{@{}c@{}}ConvTran2D,   kernel size=(4, 4), stride=(2, 2),\\    \\ input   channels=128, output channels=128, ReLU\end{tabular} \\ \cmidrule(l){2-2} 
 & \begin{tabular}[c]{@{}c@{}}Conv2d,   Kernel size=(3, 3), stride=(1, 1)\\    \\ input   channels=128, output channels=64, ReLU\end{tabular} \\ \cmidrule(l){2-2} 
 & \begin{tabular}[c]{@{}c@{}}Conv2d,   Kernel size=(3, 3), stride=(1, 1)\\    \\ input   channels=64, output channels=32, ReLU\end{tabular} \\ \cmidrule(l){2-2} 
 & \begin{tabular}[c]{@{}c@{}}Conv2d,   Kernel size=(3, 3), stride=(1, 1)\\    \\ input   channels=32, output channels=3, Sigmoid\end{tabular} \\ \bottomrule
\end{tabular}
\end{table}

\section{Experiments and Numerical Results}
In this section, a series of experiments will be implemented to evaluate the performance of image semantic transmission under the new paradigm of semantic communication system in various scenarios, and compared with several benchmarks. In particular, we consider the effects of both AWGN and fading channels on model propagation and semantic propagation at different compression rates.

We use MNIST and Cifar10 datasets to verify the feasibility of the proposed semantic image communication system. The MNIST database of handwritten digits has a training set of 60,000 examples, and a test set of 10,000 examples and the image in the MINST dataset is a single channel gray level image with a size of 28 x 28. CIFAR-10 is a subset of the 80 million tiny images dataset and consists of 60,000 32x32 color images containing one of 10 object classes, with 6000 images per class. MNIST and Cifar10 are widely used in the field of computer vision, so we think it is persuasive to use these two data sets to verify the proposed semantic communication system.

For each dataset, we design the corresponding image semantic encoder and decoder to adapt to different image sizes.  The architectures of the encoder and decoder are implemented in Pytorch. We use the Adam optimization framework, which is a form of stochastic gradient descent. The loss function of the training model is presented in formula (\ref{loss}), and its parameter $\gamma$ will be adjusted according to the characteristics of different data sets.

The performance of AESC-I and all benchmark schemes are quantified according to PSNR and SSIM, which are often used to measure image quality. Meanwhile, learned perceptual image patch similarity (LPIPS)\cite{8578166}, known as perceptual distance, is also used to compare the perceptual similarity of images. The above three indicators are measured from the image quality, lacking semantic consideration. From the perspective of semantics, referring to the semantic evaluation index proposed in formula \ref{metric}, we use a model with excellent classification ability to recognize the original image and the restored image, and the ratio of recognition rate (RR) is used as an index to judge the degree of semantic restoration.

\subsection{Evaluation on MNIST}

\subsubsection{Model structure and training}
We start by evaluating our semantic communication system on the MNIST dataset. The semantic encoder and decoder structure inspired by the autoencoder for MNIST dataset is given in Table \ref{mnist_structure}. The images of MNIST dataset are first semantically extracted through three convolution layers, then down sampled by setting the convolution stride to 2, and output the feature map with the number of channels of 16 and the length and width values of 2. After reshape operation, they are further compressed through the full connection layer. The compression ratio can be adjusted by adjusting the parameter $Z_{dims}$, given by,

\begin{equation}
Cr = \frac{C\times W \times H}{Z_{dims}},
\end{equation}
where $C$, $W$, $H$ respectively represent the number of channels, length and width of the input image.

The decoder is symmetrical with the encoder. The received semantic coding is decompressed through a full connection layer, and the feature map with size of [16,2,2] is restored through reshape operation. The feature map is processed by three-layer transpose convolution to restore the size of the source image. It is worth noting that the activation function of the last layer of transpose convolution is the Sigmoid function, which is helpful to map the image to the [0,1] interval.

When training the model, we set the learning rate and batch size to 0.001 and 256, respectively. Since the MNIST dataset is roughly a binary image dataset, we changed the MSE loss in the loss function to the BCE loss,which can speed up the convergence of the loss function. In this experiment, $\gamma = 0.50$. The model is trained until the validation loss no longer decrease.

\begin{equation}
L(S,\hat{S}) = \gamma \cdot L_{SE} + (1-\gamma) \cdot L_{BCE},
\end{equation}

\begin{equation}
\begin{aligned}
    L_{BCE}(S,\hat{S}) &= -\frac{1}{w \cdot h \cdot  c}\sum_{i=1}^{w}\sum_{j=1}^{h}\sum_{k=1}^{c}\\
    &(S_{ijk}\cdot{\log\hat S_{ijk}} + (1-S_{ijk})\cdot \log(1-\hat S_{ijk})).
\end{aligned}
\end{equation}

An example of the loss vs. epoch during the training process is shown in Fig.\ref{mnist_loss}. It can be seen from the figure that BCE loss decreases steadily and converges after 50 iterations. In the actual experiment, BCE loss has the effect of accelerating the convergence speed. The semantic loss also converges after 50 iterations, but the loss decreases unduly.

\subsubsection{Performance comparison and analysis}
According to the new paradigm of semantic communication system, the training of semantic encoder and decoder is completed at the receiver. The semantic encoder extracts and compresses the semantic coding of the source information, which is sent to the receiver together with the decoder parameters. The receiver uses the received decoder parameters to recover the source information from semantic coding. The specific structure for the image semantic communication system is shown in Fig.\ref{Image system}. LDPC channel encoding is performed for binary encoding quantified by semantic coding and decoder parameters, and the bitstream transmission is performed over a hypothetical binary discrete channel. In the simulation of communication process, we consider two widely used channel models: AWGN channel and the slow fading channel. 

Firstly, we investigate the performance of AESC-I with different compression rates in AWGN channel environment. We set the noise variance ${\sigma}^{2}$ to change the signal-to-noise ratio, and compare the proposed AESC-I with JPEG, the most widely used image compression coding technology. We can change the $Z_{dim}$ of AESC-I structure to change the setting of compression ratio, but JPEG cannot be compressed at any fixed proportion. However, for images with similar structure, JPEG compression with the same quality setting will lead to similar image compression ratio. Based on this rule, we conduct experiments on 10000 test samples of MNIST data set and set the compression ratio to 4, 8 and 20, corresponding to $Z_{dim}$ equals 196, 98, 40, corresponding to JPEG compression quality level equals 60, 18, 0. It is worth noting that part of the JPEG coding do not contain image information, but contain the necessary format information required to decoding, which is similar to the decoder parameters in AESC-I. But we do not take it into account when calculating the JPEG compression ratio and assume that the receiver knows this information, because once the format information is wrong, the image cannot be restored. Our operation undoubtedly increases the boundary condition of JPEG. 

Fig.\ref{performance_compress} illustrates the performance comparison between the AESC-I and JPEG relative to the compression ratio under different SNR conditions. Fig.\ref{performance_compress}(a) shows that the structural similarity of JPEG and AESC-I is very low in the low SNR region, and the structural similarity of  AESC-I is obviously lower than that of JPEG algorithm in the high SNR region. PSNR index in Fig.\ref{performance_compress}(b) also shows a similar situation. In the region with high SNR, the images restored by AESC-I are almost distorted from the PSNR index. However, from the LPIPS index and recognition rate ratio in Fig.\ref{performance_compress}(c) and (d), AESC-I performs better than the performance of JPEG algorithm in the region of medium and high SNR. LPIPS represents the perception distance of the neural network. The smaller the distance, the higher the perception similarity. From the perspective of the role of semantic information, the role of transmitting image semantic information on subsequent tasks can represent its semantic retention degree. The recognition rate ratio of the classification model to the received image largely represents the retention degree of its semantic information. 

In particular, Fig.\ref{performance_snr_mnist} shows the LPIPS and recognition rate ratio of different SNR in Gaussian channel with 20 times compression ratio. Direct transmission refers to no processing at the source, except quantization and LDPC coding, which is used as the upper bound of image transmission. In the region of medium and high SNR, the recognition rate ratio of AESC-I and JPEG reaches the 
upper bound of direct image transmission, but the perception distance of AESC-I is higher than JPEG. In addition, both JPEG and AESC-I have obvious cliff effect, which will deteriorate rapidly at low SNR region, but the cliff effect of AESC-I is alleviated.

The same method is used for experimental analysis of AESC-I under the slow Rayleigh fading channel with AWGN. The performance of the AESC-I and JPEG under the slow Rayleigh fading channel compared with the image direct transmission was shown in Fig.\ref{performance_snr_fading_mnist}. Similar to the performance in AWGN, cliff effect occurs in the region of low SNR. In terms of LPIPS and recognition rate ratio, the performance of AESC-I is better than JPEG. In the region of high SNR, the recognition rate ratio can reach the upper bound of direct image transmission. The perception distance of JPEG is less than that of AESC-I.

Finally, Fig.\ref{mnist_example_awgn} and Fig.\ref{mnist_example_fading} show a visual comparison of the reconstructed images of the source and channel coding schemes considered in the AWGN channel and slow fading channel. In the low SNR region, both AESC-I and JPEG failed to reconstruct the image. With the improvement of SNR, the recovery effects of the two schemes are different. JPEG realizes pixel level reconstruction, but still retains block noise. AESC-I does not retain pixel level details, but retains the complete meaning of numbers, and the reconstructed image is clear without obvious noise. This also explains why the advantages and characteristics of AESC-I can not be seen from SSIM indicators. In detail, there is a gap in the upper left corner of the  digital "0" in Fig.\ref{mnist_example_awgn}(a). The image reconstructed by JPEG algorithm retains this gap, while AESC-I restores the complete number "0", ignoring the details.  In Fig.\ref{mnist_example_awgn}(b), the circle in the lower left corner of handwritten numeral "2" is caused by personal writing habits. But the circle is erased, indicating that the AESC-I believes that it is not the semantic information of numeral 2 itself. Similar to the case in AWGN channel, in slow fading channel, the reconstruction of digital "6" and digital "8" also shows the characteristics of AESC-I semantic reconstruction. The digital "6" and "8" in Fig.\ref{mnist_example_fading} are also deformed due to personal writing habits, but AESC-I extracts its important semantic information, successfully reconstructs the image with complete digital meaning. Compared with JPEG image, the irrelevant noise is greatly reduced.

\subsection{Evaluation on Cifar-10}
\subsubsection{Model struction and training}

We also evaluate the proposed semantic communication system for image, AESC-I on more complex dataset resolution images. The specific structure and parameters of semantic encoder and decoder for Cifar-10 are shown in Table \ref{cifar_structure}. Because Cifar-10 dataset is more diverse than MNIST dataset, its design structure will be more complex. After each down sampling convolution of semantic coding structure, three convolution layers are added to further extract semantic information. The symmetric semantic decoding structure also adds two convolution layers after transpose convolution. The image from Cifar-10 dataset with the size of $[3, 32, 32]$ will be output as the feature map of $[Z_{dim}, 6, 6]$ after semantic coding. It is also necessary to restore the encoding of one-dimensional sequence to the feature map of $[Z_{dim}, 6, 6]$ before decoding. In order to reduce the size of model parameters, we no longer use the full connection layer for compression, but set $Z_{dim}$ to control the compression ratio. The compression ratio for Cifar-10 dataset can be calculated by,

\begin{equation}
Cr = \frac{C\times W \times H}{Z_{dims}\times 6\times 6} .
\end{equation}

We set the learning rate of the training model as 0.001 and the batch size as 128. The loss function is a combination of semantic loss and MSE loss, that is, formula \ref{loss}, where $\gamma$ is set to 0.1, which is used to balance the scale of the two losses. An example of the loss vs. epoch during the training process is shown in Fig.\ref{cifar_loss}. MSE loss function decreases gradually, while semantic loss decreases unsteadily, and the iteration converges after 100 iterations.

\subsubsection{Performance comparison and analysis}

The simulation flow for Cifar-10 is consistent with that for MNIST, refer to Fig.\ref{Image system}. The only difference is that the decoder parameters are quantized by 16bit instead of 8bit. Firstly, we use the idea of experiment on MNIST dataset to determine the average compression ratio under the quality level setting of JPEG with all Cifar-10 test sets, and compare it with the same compression ratio of AESC-I. We set the compression ratio to 3, 6 and 12, corresponding to $Z_{dim}$ equals 32, 12, 8, corresponding to JPEG compression quality level equals 98, 90 and 60. Similarly, the JPEG compression ratio is calculated without considering its fixed format information.

Fig.\ref{performance_compress_cifar} shows the performance of the LPIPS and recognition rate ratio of AESC-I and the JPEG algorithm under different compression ratios and different SNR. It has been verified in MNIST dataset that SSIM and PSNR can not reflect the degree of semantic recovery, so SSIM and PSNR will not be discussed in Cifar-10 anymore. Fig.\ref{performance_compress_cifar}(b) shows that the perceptual distance of the image reconstructed by AESC-I is higher than that of the image reconstructed by JPEG algorithm. However, the recognition rate ratio is similar, and the performance index of AESC-I is more stable under different compression rates. In particular, we plot the performance comparison diagram of LPIPS and recognition rate ratio with compression rate of 12 in the AWGN channel and the slow fading channel as shown in the Fig.\ref{snr_performance_awgn_cifar} and Fig.\ref{snr_performance_fading_cifar}. Cliff effect appears in both AWGN channel and the slow fading channel. In the region of low SNR, the performance of AESC-I and JPEG is almost the same. With the increase of the SNR, the performance superiority of AESC-I gradually appears. The recognition rate ratio also reaches the upper bound of image transmission. 

Fig.\ref{cifar_example_awgn} and Fig.\ref{cifar_example_fading} show some examples of image semantic transmission under different SNR of AWGN channel and the slow fading channel respectively. It is obvious that JPEG algorithm and AESC-I failed to reconstruct the image in the region of low SNR. Specifically, reconstructed images by JPEG algorithm exist in the form of block noise, and the reconstructed images by the AESC-I exist in the form of salt and pepper noise. With the increase of SNR, the reconstructed images become clearer. In particular, AESC-I can reconstruct the key parts of the image first. For example, the reconstructed images with 2.5dB noise in Fig.\ref{cifar_example_awgn}(a) and 8dB noise in Fig.\ref{cifar_example_fading}(a) have more obvious semantic features than the images reconstructed by JPEG algorithm. The background color of the reconstructed image with 3dB SNR is completely changed in Fig.\ref{cifar_example_awgn}(b), but the whole vehicle is retained. We believe that this is a manifestation of semantic feature retention.

\section{Conclusion}
In this paper, we propose a new communication paradiagm, which integrates artificial intelligence and communication, the semantic communication. A general semantic metrics is proposed from the meaning of semantics. Specifically, a semantic communication system for image, namely AESC-I is designed to verify the feasibility of the paradigm. Different from other semantic communication systems, in the semantic coding part, we not only transmit semantic coding, but also transmit semantic decoders. In addition, we use the recognition rate ratio as the performance metrics of the semantic communication for image. Experiments on MNIST and Cifar-10 datasets show that our proposed semantic communication system can effectively transmit semantic features and recover images at the receiver.




\begin{thebibliography}{10}
\providecommand{\url}[1]{#1}
\csname url@samestyle\endcsname
\providecommand{\newblock}{\relax}
\providecommand{\bibinfo}[2]{#2}
\providecommand{\BIBentrySTDinterwordspacing}{\spaceskip=0pt\relax}
\providecommand{\BIBentryALTinterwordstretchfactor}{4}
\providecommand{\BIBentryALTinterwordspacing}{\spaceskip=\fontdimen2\font plus
\BIBentryALTinterwordstretchfactor\fontdimen3\font minus
  \fontdimen4\font\relax}
\providecommand{\BIBforeignlanguage}[2]{{%
\expandafter\ifx\csname l@#1\endcsname\relax
\typeout{** WARNING: IEEEtran.bst: No hyphenation pattern has been}%
\typeout{** loaded for the language `#1'. Using the pattern for}%
\typeout{** the default language instead.}%
\else
\language=\csname l@#1\endcsname
\fi
#2}}
\providecommand{\BIBdecl}{\relax}
\BIBdecl

\bibitem{6197583}
P.~Basu, J.~Bao, M.~Dean, and J.~Hendler, ``{Preserving Quality of Information
  by Using Semantic Relationships},'' in \emph{2012 IEEE International
  Conference on Pervasive Computing and Communications Workshops}, 2012, pp.
  58--63.

\bibitem{9475174}
M.~Kountouris and N.~Pappas, ``{Semantics-Empowered Communication for Networked
  Intelligent Systems},'' \emph{IEEE Communications Magazine}, vol.~59, no.~6,
  pp. 96--102, 2021.

\bibitem{ZHANG2021}
P.~Zhang, W.~Xu, H.~Gao, K.~Niu, X.~Xu, X.~Qin, C.~Yuan, Z.~Qin, H.~Zhao,
  J.~Wei, and F.~Zhang, ``{Toward Wisdom-Evolutionary and Primitive-Concise
  6G:A New Paradigm of Semantic Communication Networks},'' \emph{Engineering},
  2021.

\bibitem{mikolov2013distributed}
T.~Mikolov, I.~Sutskever, K.~Chen, G.~S. Corrado, and J.~Dean, ``{Distributed
  Representations of Words and Phrases and their Compositionality},'' in
  \emph{Advances in Neural Information Processing Systems}, 2013, pp.
  3111--3119.

\bibitem{devlin2018bert}
J.~D. M.-W.~C. Kenton and L.~K. Toutanova, ``{Bert: Pre-training of Deep
  Bidirectional Transformers for Language Understanding},'' in
  \emph{Proceedings of NAACL-HLT}, 2019, pp. 4171--4186.

\bibitem{mccann2017learned}
B.~McCann, J.~Bradbury, C.~Xiong, and R.~Socher, ``{Learned in Translation:
  Contextualized Word Vectors},'' in \emph{Proceedings of the 31st
  International Conference on Neural Information Processing Systems}, 2017, pp.
  6297--6308.

\bibitem{sarzynska2021detecting}
J.~Sarzynska-Wawer, A.~Wawer, A.~Pawlak, J.~Szymanowska, I.~Stefaniak,
  M.~Jarkiewicz, and L.~Okruszek, ``Detecting formal thought disorder by deep
  contextualized word representations,'' \emph{Psychiatry Research}, vol. 304,
  p. 114135, 2021.

\bibitem{he2016deep}
K.~He, X.~Zhang, S.~Ren, and J.~Sun, ``{Deep Residual Learning for Image
  Recognition},'' in \emph{Proceedings of the IEEE conference on computer
  vision and pattern recognition}, 2016, pp. 770--778.

\bibitem{ren2015faster}
S.~Ren, K.~He, R.~Girshick, and J.~Sun, ``{Faster r-cnn: Towards Real-time
  Object Detection with Region Proposal Networks},'' \emph{Advances in neural
  information processing systems}, vol.~28, pp. 91--99, 2015.

\bibitem{isensee2019nnu}
F.~Isensee, J.~Petersen, S.~A. Kohl, P.~F. J{\"a}ger, and K.~H. Maier-Hein,
  ``{nnU-Net: Breaking the Spell on Successful Medical Image Segmentation},''
  \emph{arXiv preprint arXiv:1904.08128}, vol.~1, pp. 1--8, 2019.

\bibitem{goodfellow2014generative}
I.~J. Goodfellow, J.~Pouget-Abadie, M.~Mirza, B.~Xu, D.~Warde-Farley, S.~Ozair,
  A.~Courville, and Y.~Bengio, ``{Generative Adversarial Nets},'' \emph{stat},
  vol. 1050, pp. 10--19, 2014.

\bibitem{karras2019style}
T.~Karras, S.~Laine, and T.~Aila, ``{A style-based Generator Architecture for
  Generative Adversarial Networks},'' in \emph{Proceedings of the IEEE/CVF
  Conference on Computer Vision and Pattern Recognition}, 2019, pp. 4401--4410.

\bibitem{dahl2011context}
G.~E. Dahl, D.~Yu, L.~Deng, and A.~Acero, ``{Context-dependent Pre-trained Deep
  Neural Networks for Large-vocabulary Speech Recognition},'' \emph{IEEE
  Transactions on audio, speech, and language processing}, vol.~20, no.~1, pp.
  30--42, 2011.

\bibitem{o2016learning}
T.~J. O'Shea, K.~Karra, and T.~C. Clancy, ``{Learning to Communicate: Channel
  Auto-encoders, Domain Specific Regularizers, and Attention},'' in \emph{2016
  IEEE International Symposium on Signal Processing and Information Technology
  (ISSPIT)}, 2016, pp. 223--228.

\bibitem{xie2021deep}
H.~Xie, Z.~Qin, G.~Y. Li, and B.-H. Juang, ``{Deep Learning Enabled Semantic
  Communication Systems},'' \emph{IEEE Transactions on Signal Processing},
  vol.~69, pp. 2663--2675, 2021.

\bibitem{xie2020lite}
H.~Xie and Z.~Qin, ``{A Lite Distributed Semantic Communication System for
  Internet of Things},'' \emph{IEEE Journal on Selected Areas in
  Communications}, vol.~39, no.~1, pp. 142--153, 2020.

\bibitem{weng2021semantic}
Z.~Weng and Z.~Qin, ``{Semantic Communication Systems for Speech
  Transmission},'' \emph{IEEE Journal on Selected Areas in Communications},
  vol.~39, no.~8, pp. 2434--2444, 2021.

\bibitem{9066966}
D.~B. Kurka and D.~Gündüz, ``{DeepJSCC-f: Deep Joint Source-Channel Coding of
  Images With Feedback},'' \emph{IEEE Journal on Selected Areas in Information
  Theory}, vol.~1, no.~1, pp. 178--193, 2020.

\bibitem{9154306}
M.~Jankowski, D.~Gündüz, and K.~Mikolajczyk, ``{Joint Device-Edge Inference
  over Wireless Links with Pruning},'' in \emph{{2020 IEEE 21st International
  Workshop on Signal Processing Advances in Wireless Communications (SPAWC)}},
  2020, pp. 1--5.

\bibitem{8723589}
E.~Bourtsoulatze, D.~Burth~Kurka, and D.~Gündüz, ``{Deep Joint Source-Channel
  Coding for Wireless Image Transmission},'' \emph{IEEE Transactions on
  Cognitive Communications and Networking}, vol.~5, no.~3, pp. 567--579, 2019.

\bibitem{8578166}
R.~Zhang, P.~Isola, A.~A. Efros, E.~Shechtman, and O.~Wang, ``{The Unreasonable
  Effectiveness of Deep Features as a Perceptual Metric},'' in \emph{2018
  IEEE/CVF Conference on Computer Vision and Pattern Recognition}, 2018, pp.
  586--595.

\end{thebibliography}
\end{document}